\pdfoutput=1
\documentclass[10pt,twocolumn,a4paper,aps,prd,preprintnumbers,showpacs,superscriptaddress,nofootinbib,amsmath,amssymb]{revtex4-1}
\usepackage{cases}
\usepackage[utf8]{inputenc}
\usepackage[T1]{fontenc}
\usepackage{cmap}

\newcommand{\newc}{\newcommand}
\newc{\beq}{\begin{equation}}
\newc{\eeq}{\end{equation}}
\newc{\bal}{\begin{align}}
\newc{\eal}{\end{align}}
\newc{\ba}{\begin{eqnarray}}
\newc{\ea}{\end{eqnarray}}
\newc{\bea}{\begin{eqnarray*}}
\newc{\eea}{\end{eqnarray*}}
\newc{\alp}{\alpha}
\newc{\eps}{\epsilon}
\newc{\vph}{\varphi}
\newc{\vhp}{\varphi}

\usepackage{graphicx,rotating,colortbl}

\definecolor{nicered}{rgb}{0.7,0.1,0.1}
\definecolor{nicegreen}{rgb}{0.1,0.5,0.1}
\definecolor{rosso}{cmyk}{0,1,1,0.4}
\definecolor{babypink}{rgb}{0.96, 0.76, 0.76}
\definecolor{babyblueeyes}{rgb}{0.63, 0.79, 0.95}
\definecolor{azure(colorwheel)}{rgb}{0.0, 0.5, 1.0}
\definecolor{amethyst}{rgb}{0.6, 0.4, 0.8}
\definecolor{MyDarkBlue}{rgb}{0,0.1,0.7}
\definecolor{secnum}{RGB}{13,151,225}
\definecolor{ptcbackground}{RGB}{212,237,252}
\definecolor{ptctitle}{RGB}{0,177,235}
\definecolor{blus}{cmyk}{1,1,0,0.1}
\definecolor{verdes}{cmyk}{0.99,0,0.59,0.65}
\definecolor{rossos}{cmyk}{0,1,1,0.55}
\definecolor{redy}{cmyk}{0,1,1,0.7}
\definecolor{greeny}{cmyk}{0.99,0,0.59,0.98}
\definecolor{green-go}{cmyk}{0.79,0,0.59,0.5}
\definecolor{Maroon}{RGB}{128,0,0}
\definecolor{purple}{RGB}{153,0,153}
\definecolor{DarkViolet}{RGB}{148,0,211}

\usepackage{hyperref}
\hypersetup{colorlinks,bookmarksopen,
bookmarksnumbered,
citecolor={DarkViolet},
linkcolor={DarkViolet},
urlcolor={DarkViolet},
pdfstartview=FitH} 
\graphicspath{{./Figs/}}

\begin{document}

\title{Einstein-scalar-Gauss-Bonnet black holes: Analytical approximation for the metric and applications to calculations of shadows}

\author{Roman A. Konoplya}\email{roman.konoplya@gmail.com}
\affiliation{Institute of Physics and Research Centre of Theoretical Physics and Astrophysics, Faculty of Philosophy and Science, Silesian University in Opava, Bezručovo nám. 13, CZ-746 01 Opava, Czech Republic}
\affiliation{Peoples Friendship University of Russia (RUDN University), 6 Miklukho-Maklaya Street, Moscow 117198, Russian Federation}

\author{Thomas Pappas}
\email{thomasdpappas@gmail.com}
\affiliation{Institute of Physics and Research Centre of Theoretical Physics and Astrophysics, Faculty of Philosophy and Science, Silesian University in Opava, Bezručovo nám. 13, CZ-746 01 Opava, Czech Republic}
\affiliation{Division of Theoretical Physics, Department of Physics, University of Ioannina, Ioannina GR-45110, Greece}

\author{Alexander Zhidenko}\email{olexandr.zhydenko@ufabc.edu.br}
\affiliation{Institute of Physics and Research Centre of Theoretical Physics and Astrophysics, Faculty of Philosophy and Science, Silesian University in Opava, Bezručovo nám. 13, CZ-746 01 Opava, Czech Republic}
\affiliation{Centro de Matemática, Computação e Cognição (CMCC), Universidade Federal do ABC (UFABC),\\ Rua Abolição, CEP: 09210-180, Santo André, SP, Brazil}

\begin{abstract}
Recently, numerical solutions to the field equations of Einstein-scalar-Gauss-Bonnet gravity that correspond to black-holes with non-trivial scalar hair have been reported. Here, we employ the method of the continued-fraction expansion in terms of a compact coordinate in order to obtain an analytical approximation for the aforementioned solutions. For a wide variety of coupling functionals to the Gauss-Bonnet term we were able to obtain analytical expressions for the metric functions and the scalar field. In addition we estimated the accuracy of these approximations by calculating the black-hole shadows for such black holes. Excellent agreement between the numerical solutions and analytical approximations has been found.
\end{abstract}


\date{\today}

\maketitle

\section{Introduction}
\label{intro}

Nowadays black holes are the most important objects for understanding the regime of strong gravity. Observations in the gravitational \cite{abbott2016,abbott2016a,abbott2016b} and electromagnetic \cite{Akiyama:2019cqa,Goddi:2017pfy} spectra support General Relativity, but, at the same time, leave ample room for alternative theories of gravity \cite{Yunes:2016jcc,Konoplya:2016pmh}. One of the most interesting alternative approaches is related to adding higher-curvature corrections to the Einstein action. This kind of extension of the Einstein gravity is inspired by the low-energy limit of string theory \cite{Callan:1988hs,Kanti1996} and, presumably, could describe quantum corrected black holes. The lowest-order correction is given by the (second order in curvature) Gauss-Bonnet term, which is pure divergence in four-dimensional spacetimes, but, when coupled to a scalar field, it leads to modifications of the Einstein equations.

All the known black-hole solutions in the four-dimensional Einstein-scalar-Gauss-Bonnet gravity are obtained either numerically \cite{Kanti1996,Antoniou2018,Kleihaus:2015aje,Collodel:2019kkx,Cunha:2016wzk,Blazquez-Salcedo:2017txk}, or perturbatively \cite{Ayzenberg:2014aka,Ayzenberg:2014aka,Maselli:2015tta}, which makes the usage of a number of tools for analysis of behavior of such solutions either difficult or impossible. Analytical expressions for such numerical black-hole metrics, which are valid in the whole space outside the event horizon, would allow us to see the explicit dependence of the metric on physical parameters of the system and to work with the metric as, essentially, an exact solution. The approach to finding analytical approximations of numerical solutions was based on the general parametrization for spacetimes of static spherically symmetric black holes \cite{Rezzolla2014} and extended in \cite{Konoplya:2016jvv} to axial symmetry.
For spherical symmetry the parametrization uses a continued-fraction expansion in terms of a compactified radial coordinate. This choice leads to superior convergence properties and allows one to approximate a black-hole metric with a much smaller set of coefficients. This approach was used to construct the analytical approximation of numerical black-hole solutions in the Einstein-Weyl \cite{Kokkotas2017a}, Einstein-dilaton-Gauss-Bonnet \cite{Kokkotas2017} and Einstein-scalar-Maxwell \cite{Konoplya:2019goy} theories. Further studies of observables in these parametrized spacetimes \cite{Younsi:2016azx,Konoplya:2018arm,Nampalliwar:2018iru,Konoplya:2019hml,Konoplya:2019ppy,Zinhailo:2018ska,Zinhailo:2019rwd} showed that usually only 2 to 3 orders of the continued-fraction expansion are sufficient in order to achieve reasonable accuracy.

In \cite{Kokkotas2017} the analytical approximation was found for the particular choice of the scalar field coupling functional -- the dilaton, exponential coupling, which was considered numerically in  \cite{Kanti1996}. Recently this approach was extended in  \cite{Antoniou2018} to various types of the scalar-field functional and, therefore, allowed one to look whether there are some common features for all the considered couplings of the scalar field. A similar problem was attacked numerically for the case of the Einstein-scalar-Maxwell theory \cite{Herdeiro:2018wub} and the study of its analytical approximation  \cite{Konoplya:2019goy} showed that the radius of the black-hole shadow is increased for any of the considered couplings of the scalar field. \emph{Scalarization}, that is, the phenomenon of spontaneous acquiring of a scalar hair by the black hole as a result of the nonminimal coupling of a scalar field to the system, has been actively studied in \cite{Doneva:2017bvd,Silva:2017uqg,Minamitsuji:2018xde,Cunha:2019dwb,Fernandes:2019rez}.

Here we generalize the procedure for finding the analytical approximation in the Einstein-scalar-Gauss-Bonnet (EsGB) theory to the cases of various coupling functionals of the scalar field. Then, we apply the obtained parametrized black-hole metrics to the calculation of the radii of shadows in order to estimate the relative error due to the truncation of the continued-fraction expansion which we used. We also present the analytical expressions for both the radius of the photon sphere and the black-hole shadow.

The paper is organized as follows. In Sec.~\ref{sec:ESGB} we present the basics of the EsGB theory. Sec.~\ref{sec:CF} is devoted to the introduction of the continued-fraction expansion, while in Sec.~\ref{sec:AAESGB} we apply this procedure to the numerical solution of the EsGB black holes. Finally in Sec.~\ref{sec:shadows} we find black-hole shadows for the above numerical and parametrized black-hole metrics. In the Conclusions we summarize the obtained results and discuss the open questions.

\section{Black holes in Einstein-scalar-Gauss-Bonnet gravity}
\label{sec:ESGB}
The Lagrangian for EsGB gravity reads
\beq
\mathcal{L}= \frac{R}{\kappa^2}+\alp f(\vph) R^2_{GB}-\frac{1}{2} \nabla_{\rho}\vph \nabla^{\rho} \vph \,,
\label{action}
\eeq
where $\kappa^2 \equiv 16 \pi G c^{-4}=1$ is the  Einstein's constant. The Gauss-Bonnet (GB) term is defined as
\beq
R^2_{GB} \equiv \left( R^2 - 4 R_{\beta\gamma}R^{\beta\gamma}+R_{\beta\gamma\rho\sigma} R^{\beta\gamma\rho\sigma} \right)\,,
\eeq
while $\alp$ is the GB constant and $f(\varphi)$ is an arbitrary smooth function of the scalar field $\vph$ corresponding to GB-coupling functional.

In four dimensions, if $f(\vph)$ is a constant, then the GB term is \textit{topological} in the sense that it does not contribute to the field equations. In the case of an exponential coupling functional $f(\vph) = e^{\vph}$ black-hole solutions with scalar hair emerge for EsGB gravity and the first solutions were obtained numerically in \cite{Kanti1996}. More recently, the authors of \citep{Antoniou2018a} have reported that regular black-hole solutions with scalar hair appear as a generic feature of the theory \eqref{action}.

Let us start by considering the following line element for a static and spherically symmetric spacetime:
\beq
ds^2=-g_{tt}(r) dt^2+g_{rr}(r) dr^2+r^2 \left( d \theta ^2 +\sin ^2\theta d \phi ^2 \right) \,.
\label{metric ansatz}
\eeq
We also assume that the scalar field shares the symmetries of the underlying spacetime and it thus depends solely on the radial coordinate $r$.

The Einstein equations that are derived from the theory \eqref{action} are the following:

\beq\label{Tmn}
\begin{split}
&R_{\mu\nu}-\frac{1}{2} R \,g_{\mu\nu}=-\frac{1}{4}g_{\mu\nu}\partial_{\rho}\vph\partial^{\rho}\phi+\frac{1}{2}\partial_{\mu}\phi\partial_{\nu}\vph
\\&
-\frac{1}{2}\left(g_{\rho\mu}g_{\lambda\nu}+g_{\lambda\mu}g_{\rho\nu}\right)
\eta^{\eta\lambda\sigma\beta}\tilde{R}^{\rho\gamma}_{\quad\sigma\beta}
\nabla_{\gamma}\partial_{\eta}\left[\alp f(\vph)\right] \,,
\end{split}
\eeq

where
\beq\label{g}
\tilde{R}^{\rho\gamma}_{\quad\alpha\beta}=\eta^{\rho\gamma\sigma\tau}
R_{\sigma\tau\alpha\beta}=\frac{\epsilon^{\rho\gamma\sigma\tau}}{\sqrt{-g}}\,
R_{\sigma\tau\alpha\beta}\,,
\eeq
and the scalar-field equation of motion is
\beq
\nabla^2 \vph+\alp f'(\vph)R^2_{GB}=0\,, \label{phi-eq_0}
\eeq

where it is understood that throughout this article a prime indicates differentiation with respect to the argument of the function.

Numerical solutions to the field equations of EsGB gravity corresponding to black holes with scalar hair have been recently found in \cite{Antoniou2018} for a wide range of GB couplings. Here, by employing the method of \cite{Rezzolla2014} we obtain analytical approximations of these numerical solutions.

\section{The continued-fraction approximation}\label{sec:CF}
In this section we outline the method of the continued-fraction approximation (CFA) \cite{Rezzolla2014} and introduce the notations we use in the rest of the article.

In the original coordinate system of \eqref{metric ansatz}, the radius of the event horizon of the black hole $r_0$ is determined by the vanishing of the norm of the timelike Killing vector associated with the invariance of the metric under time translations. This condition eventually translates to $g_{tt}(r_0)=0$. Then, we may perform a radial coordinate transformation and introduce the compact coordinate
\beq
x \equiv 1-\frac{r_0}{r}\,,
\eeq
that ranges from $x=0$ at the location of the horizon up to $x=1$ at spatial infinity.

In the CFA, we consider a new metric ansatz that is suitable for approximating any spherically symmetric metric to high accuracy with only a small number of parameters \cite{Kokkotas2017a,Kokkotas2017}. The metric coefficients of \eqref{metric ansatz} are written in terms of the new set of functions $A(x)$ and $B(x)$ defined via the following relations:
\beq
\begin{split}
& g_{tt}(r)= x A(x)\,,\\
& g_{tt}(r)g_{rr}(r)= B(x)^2\,,
\end{split}
\label{A(x) and B(x) defs}
\eeq
with
\beq
\begin{split}
A(x)&\equiv 1-\eps (1-x)+(a_0-\eps)(1-x)^2
+\tilde{A}(x)(1-x)^3 \\
B(x)&\equiv 1+b_0 (1-x) + \tilde{B}(x)(1-x)^2\,,
\end{split}
\label{RZ ansatz}
\eeq
where the parameter $\eps$ is determined by the value of the  asymptotic mass $M$ of the black hole and the location of its event horizon $r_0$ as
\beq
\eps \equiv -\left(1-\frac{2 M}{r_0} \right)\,.
\label{eps definition}
\eeq
The parameter $\eps$ indicates the amount of the deviation of the EsGB black-hole geometry from the Schwarzschild black hole, for which $r_0 = 2\,M $. The parameters $a_0$ and $b_0$ are defined in terms of $\eps$ and the so-called parametrized post-Newtonian parameters $\beta$ and $\gamma$ as
\begin{eqnarray}
a_0 &\equiv& \frac{(\beta -\gamma)(1+\epsilon)^2}{2},\\
b_0 &\equiv& \frac{(\gamma-1)(1+\epsilon)^2}{2}.
\end{eqnarray}

The functions ${\tilde A}(x)$ and ${\tilde B}(x)$ have the delicate role of describing the metric near the horizon ($x=0$) and are defined in terms of continued-fraction expansions as follows:

\begin{align}\nonumber
{\tilde A}(x)=\frac{a_1}{\displaystyle 1+\frac{\displaystyle
    a_2x}{\displaystyle 1+\frac{\displaystyle a_3x}{\displaystyle
      1+\frac{\displaystyle a_4x}{\displaystyle
      1+\ldots}}}}\,,\\\label{contfrac}
{\tilde B}(x)=\frac{b_1}{\displaystyle 1+\frac{\displaystyle
    b_2x}{\displaystyle 1+\frac{\displaystyle b_3x}{\displaystyle
      1+\frac{\displaystyle b_4x}{\displaystyle
      1+\ldots}}}}\,.
\end{align}

The values of the parameters $a_i$ and $b_i$ for $i\geqslant 1$ can be obtained numerically upon expanding both sides of Eqs.\eqref{A(x) and B(x) defs} near the horizon and comparing coefficients of the same order in the expansion.

At this point let us mention that at spatial infinity the metric functions and the scalar field can be approximated as \citep{Antoniou2018a}
\beq
\begin{split}
g_{tt}(r)=&1-\frac{2M}{r}+\frac{M D^2}{12 r^3}+\frac{24 M D f'+M^2 D^2}{6 r^4}
\\&
+\mathcal{O}(1/r^5)\,,
\label{gtt infinity}
\end{split}
\eeq

\beq
\begin{split}
g_{rr}(r)=&1+\frac{2M}{r}+\frac{16  M^2 - D^2}{4 r^2}
+\frac{32 M^3-5 M D^2}{4 r^3}
\\&
+\frac{768 M^4 -208 M^2 D^2-384 M D f'+3 D^4}{48 r^4}
\\&
+\mathcal{O}(1/r^5)\,,
\label{grr infinity}
\end{split}
\eeq

\beq
\begin{split}
\vph(r)=&\, \vph_{\infty} +\frac{D}{r}+\frac{M D}{r^2}+\frac{32 M^2 D-D^3}{24 r^3}
\\&
+\frac{12 M^3 D - 24 M^2 f' -M D^3}{6 r^4}+\mathcal{O}(1/r^5)\,,
\label{phi infinity}
\end{split}
\eeq
where $\vph_{\infty}$ is the asymptotic value of the scalar field and $D$ is its charge. Notice that the exact form of $f(\vph)$ plays no role in the asymptotic expansions up to the third order. The form of Eqs.~\eqref{gtt infinity} and \eqref{grr infinity} implies that $\beta=\gamma=1$ and thus $a_0=b_0=0$ for any GB-coupling functional.

In the same spirit, an analytical approximation for the scalar field can also be obtained by means of the CFA \cite{Kokkotas2017}. For this purpose we define a new function of the compact coordinate that is related to the scalar field and its asymptotic value at spatial infinity via the following relation:
\beq\label{Fx}
F(x) = e^{\vph(r)-\vhp_{\infty}}\,,
\eeq
where the left-hand side is expanded as
\beq
F(x)=1+f_0(1-x)+{\tilde F}(x)(1-x)^2\,.
\label{F(x) def}
\eeq
The coefficient $f_0=D/r_0$ is determined by the value of the charge of the scalar field and
\beq
{\tilde F}(x)=\frac{f_1}{\displaystyle 1+\frac{\displaystyle
    f_2x}{\displaystyle 1+\frac{\displaystyle f_3x}{\displaystyle
      1+\ldots}}}\,.
\label{F(x) CFA}
\eeq
Again, by expanding \eqref{Fx} near the event horizon one can obtain numerically the values of the coefficients $f_i$ for $i \geqslant 1$.

\section{Analytical approximations for EsGB black holes}\label{sec:AAESGB}

By employing the method described in the previous section we have derived analytical approximations for numerical black-hole solutions emerging in EsGB gravity. More precisely, for all the numerical solutions obtained for the different coupling functionals studied in \cite{Antoniou2018} we give here the approximate analytic metric coefficients.

Near the location of the event horizon we may expand the metric functions and the scalar field as follows:

\ba
g_{tt}(r) &=& p_1 \left[ (r-r_0)+\sum_{n=2}^{\infty} p_n (r-r_0)^n \right]\,,
\label{metric horizon expansion} \\\nonumber
g_{rr}(r) &=& \sum_{n=1}^{\infty} q_n (r-r_0)^n\,,
\ea
and
\beq
\vph(r)=\sum_{n=0}^{\infty} \frac{\vph_n}{n!} (r-r_0)^n \,,
\label{phi horizon expansion}
\eeq
where $\vph_n \equiv \vph^{(n)}(r_0)$ is the $n$-th order derivative of the scalar field evaluated on the event horizon.

The value of the scalar field on the horizon $\vph_0$ is a free parameter, subject to the requirement $\vph_1 \in \mathbb{R}$ in order for a black-hole solution to exist.
Upon specifying the form of the coupling functional $f(\vph)$, the first derivative of the scalar field $\vph _1$ on the horizon is uniquely determined for each value of $\vph_0$ through the constraint \cite{Antoniou2018}
\beq
\vph_1 =\frac{r_0}{4\alp f'(\vph_0)}\left( \sqrt{1-\frac{96\alp^2 f'(\vph_0)^2}{r_0^4}}-1 \right) \,.
\label{phi 1}
\eeq
Once the values of the parameters $p_1,\vph_0,$ and $\vph_1$ have been specified, the rest of the parameters $p_i,q_i$ and $\vph_i$ in the expansions can be determined recursively up to an arbitrary order. This is achieved by substituting Eqs.~\eqref{metric horizon expansion}~and~\eqref{phi horizon expansion} into the field equations and then solving the corresponding equations order by order in the expansion.

It is convenient to introduce a new dimensionless parameter $p$ instead of $\vph_0$ to parametrize the family of black-hole solutions for each GB coupling as follows:
\beq
p \equiv \frac{96 \alp^2 f'(\vph_0)^2}{r_0^4}\,.
\label{dimensionless param}
\eeq
Notice that in order to have a regular black-hole solution with a well-defined horizon \cite{Kanti1996,Antoniou2018}, the following constraint must hold via Eq.~\eqref{phi 1}:
\beq
p \in [0,1]\,,
\eeq
with the Schwarzschild limit corresponding to $p = 0$, while for $p \rightarrow 1$ the maximal-coupling regime is approached.

After fixing the units of length in such a way that $r_0=1$, $p$ depends on two parameters, $\alp$ and $\vhp_0$. In this paper we consider $\alp=1/4$ and collect numerical data for the family of black-hole solutions by varying $\vph_0$. Further, by comparing the numerical solutions for other values of $\alp$ we see that, for a fixed value of $p$ and varying $\alpha$, the variation of the black-hole geometry is negligibly small; i.e. for practical purposes we need to take into account only the value of $p$.

For each value of $p$ we numerically integrate the field equations to obtain the accurate numerical solutions for the metric functions and the scalar field\footnote{The interested reader can find more details about the numerical black-hole solutions emerging in EsGB gravity in \citep{Antoniou2018}.}. The parameter $p_1$ is then fine-tuned such that for $r \rightarrow \infty$ we have $g_{tt}(r) \rightarrow 1$ and $g_{rr}(r) \rightarrow 1$ and this way recover the asymptotically flat limit.

With these solutions at hand, the next step is to determine the values of the asymptotic parameters of the system. The asymptotic mass $M$ is computed by expanding the solution for $g_{tt}(r)$ at large values of the radial coordinate and isolating the numerical coefficient of the term $\sim 1/r$. Then, according to \eqref{gtt infinity}, $M$ simply corresponds to $-1/2 \times$(value of coefficient). This also determines the value of the parameter $\eps$ via \eqref{eps definition}. Similarly, the asymptotic value for $D$ of the scalar-field expansion \eqref{phi infinity} is determined via the corresponding coefficient of the expansion of the numerical solution for $\vph(r)$.

The numerical values for the parameters $(p_i,q_i,\vph_i)$ are thus determined as described above for each value of $\vph_0$ and $p_1$ and in this way through Eqs.\eqref{A(x) and B(x) defs} and \eqref{Fx} one finally ends up with numerical values for the set $(a_i,b_i,f_i)$.

The above steps are repeated for different values of $p$ that span the allowed range $[0,1]$ and numerical data are assembled for $(a_i,b_i,f_i)$. Then, one is able to perform a fitting of these data in order to obtain analytical expressions for the CFA parameters as functions of $p$. It is then straightforward to write down approximate analytical expressions for the metric functions and the scalar field to the desired order in the CFA via \eqref{RZ ansatz} and \eqref{F(x) def}.

\subsection{The even-polynomial coupling functional}
The first case we study is the even-polynomial coupling functional
\beq
f(\vph)= \vph^{2 n},\quad n \in \mathbb{Z}^{+}\,.
\label{coupling: even polynomial}
\eeq

The form of the dimensionless parameter \eqref{dimensionless param} for this family of black-hole solutions ($\alp=1/4$) is
\beq
p = \left(24 n^2\right) \varphi_0 ^{4 n-2}\,,
\label{p even pol}
\eeq
and the allowed values for $\vph_0$ are thus
\beq
|\vph_0 | \leqslant \left(24 n^2\right)^{\frac{1}{2-4 n}}\,.
\label{even allowed phi0s}
\eeq
In order to be able to perform the analysis we need to further reduce the number of free parameters and so we must also choose a specific value for $n$ in \eqref{coupling: even polynomial}.

As illustrative cases for this family of functionals we study $n=1$ and $n=2$ that correspond to the quadratic and quartic couplings, respectively.

\subsubsection{The quadratic GB-coupling functional}

The obtained analytical expressions for the parameters of the CFA \eqref{RZ ansatz},\eqref{contfrac},\eqref{F(x) def} and \eqref{F(x) CFA} up to second order are given below

\beq
a_1= \frac{\frac{63 p}{332}-\frac{23 p^2}{143}}{-\frac{83 p^2}{401}+p-\frac{223}{259}}\,,
\label{a1 even}
\eeq

\beq
a_2= \frac{-\frac{234 p^2}{307}+\frac{152 p}{397}+1}{\frac{73}{221}-\frac{73 p}{228}}\,,
\label{a2 even}
\eeq

\beq
\eps= \frac{\frac{p}{43}-\frac{p^2}{201}}{1-\frac{105 p}{577}}\,.
\label{eps even}
\eeq

\begin{figure}
\includegraphics[width=\linewidth]{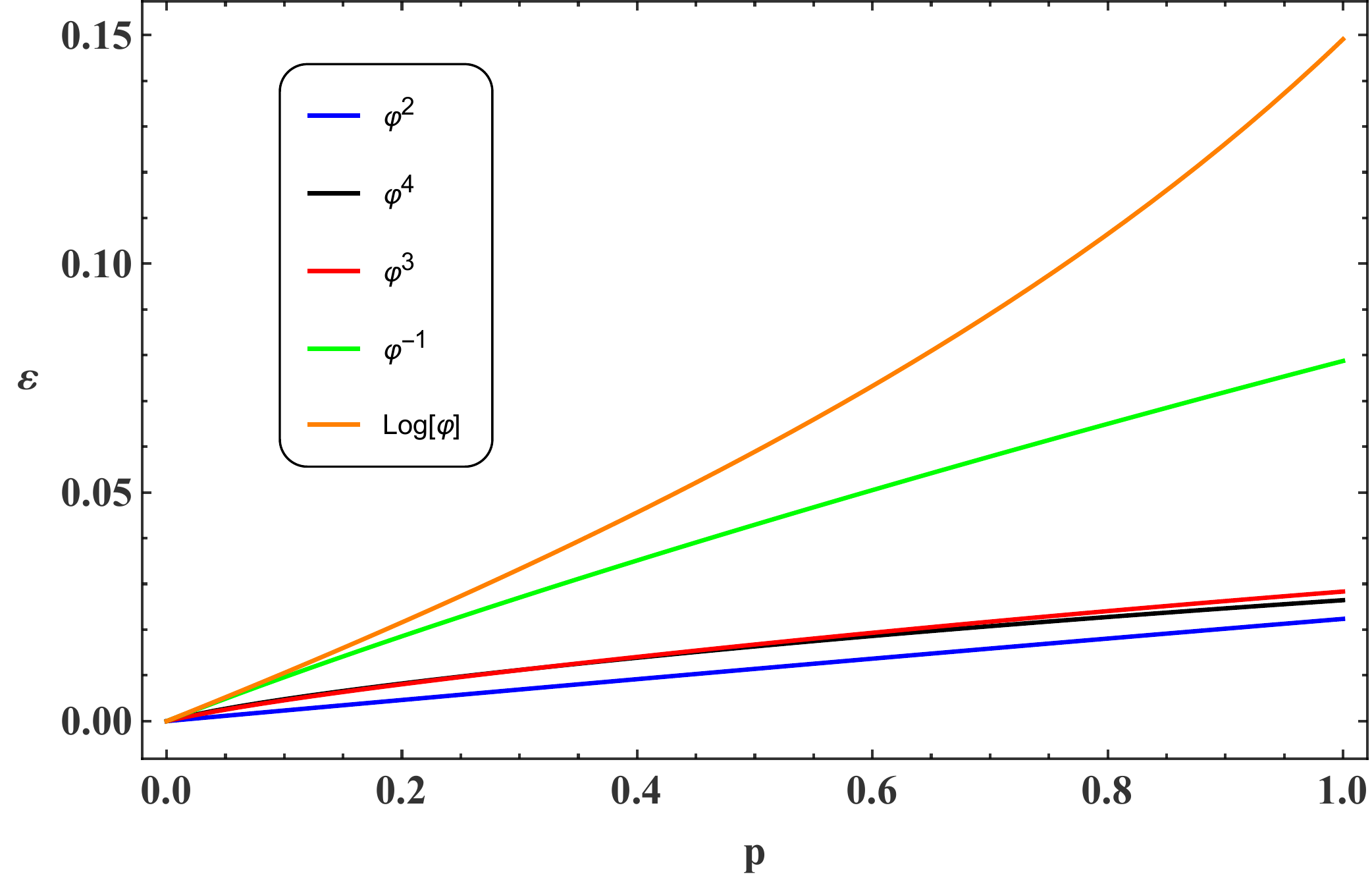}
\caption{The asymptotic parameter $\eps$ \eqref{eps definition} as a function of the dimensionless parameter p \eqref{dimensionless param} for different GB-coupling functionals. }
  \label{fig: epsilons_vs_p}
\end{figure}

The profile of the $\eps$ parameter with respect to $p$ is depicted in Fig.~\ref{fig: epsilons_vs_p} for all the GB couplings we have studied in this article.

\begin{figure}
\includegraphics[width=\linewidth]{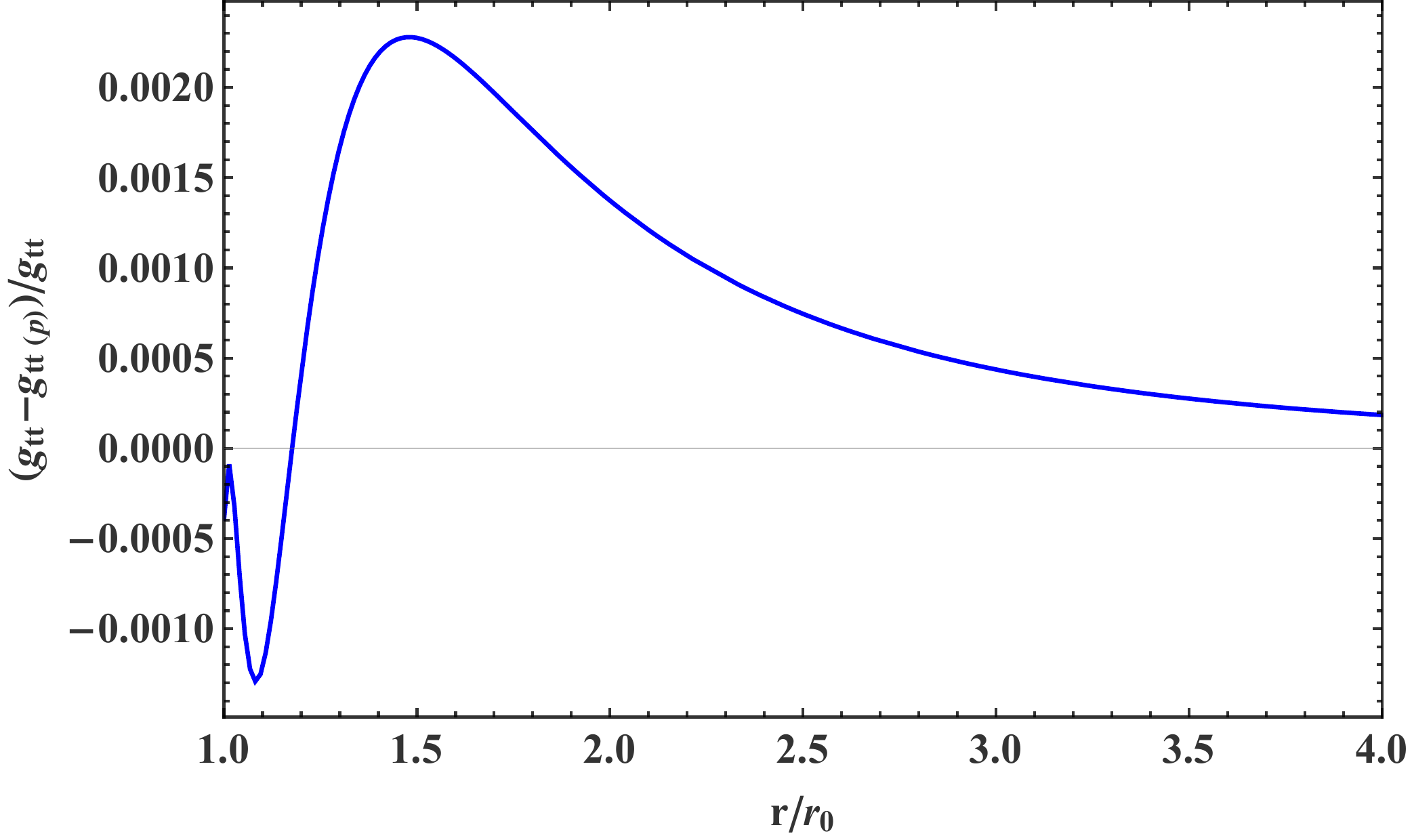}
\caption{The relative error of the fourth-order analytical approximation for $g_{tt}(r)_{(p)}$ from the accurate numerical solution $g_{tt}(r)$ for $p=0.8$. }
  \label{fig: RE_gtt_Eve_08}
\end{figure}

The above parameters alone suffice for the determination of the analytical representation of $g_{tt}(r)$. We point out that a general feature of the approximate expressions for both the metric functions and the scalar field is that the relative error (RE) increases with $p$. For the GB coupling $f(\vph)=\vph^2$ when $p=0.8$ in Fig.~\ref{fig: RE_gtt_Eve_08} we plot the RE between the fourth-order analytical approximation for the $g_{tt}(r)$ metric function and its accurate numerical solution. The maximum error occurs around the photon sphere radius at $r \approx 1.5 \,r_0$ and is less than $0.24\%$.

In turn, the analytical approximation of $g_{rr}(r)$ emerges via \eqref{A(x) and B(x) defs} and thus requires also the expressions for the parameters $b_i$ that are listed below

\beq
b_1=\frac{\frac{91 p}{396}-\frac{85 p^2}{438}}{-\frac{24 p^2}{131}+p-\frac{632}{707}}\,,
\label{b1 even}
\eeq

\beq
b_2=\frac{-\frac{173 p^3}{432}+\frac{47 p^2}{225}-\frac{106 p}{203}+1}{\frac{20}{221}-\frac{25 p}{283}}\,,
\label{b2 even}
\eeq

Finally for the scalar field the analytically-approximated parameters for the CFA are found to be

\beq
\vph_{\infty}=\frac{-\frac{9 p^2}{56}-\frac{p}{67}}{p^2+\frac{38 p}{61}+\frac{1}{145}}\,,
\label{phiInf even}
\eeq

\beq
f_0 = \frac{-\frac{11 p^3}{138}+\frac{3 p^2}{28}+\frac{p}{113}}{-\frac{47 p^3}{49}+p^2+\frac{84 p}{131}+\frac{1}{155}} \,,
\label{f0 even}
\eeq

\beq
f_1 = \frac{\frac{63 p^2}{151}+\frac{3 p}{46}}{\frac{113 p^2}{212}+p+\frac{2}{111}} \,,
\label{f1 even}
\eeq

\beq
f_2 = \frac{-\frac{23 p^3}{55}-\frac{30 p^2}{97}+p}{\frac{3 p^3}{56}-\frac{31 p^2}{79}+\frac{79 p}{211}+\frac{1}{8997}} \,,
\label{f2 even}
\eeq

\begin{figure*}
\includegraphics[angle=0.0,width=0.5\linewidth]{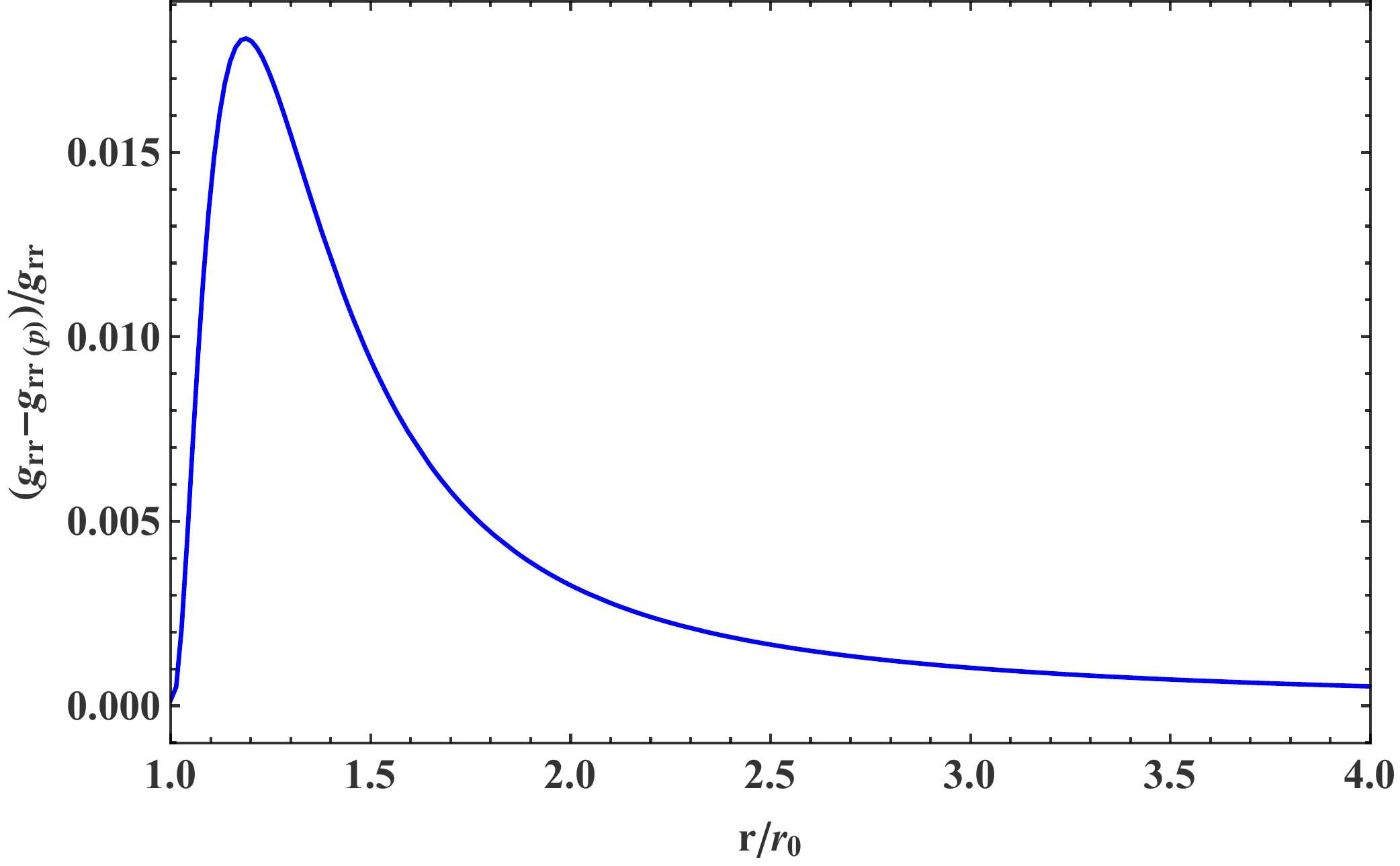}\includegraphics[angle=0.0,width=0.5\linewidth]{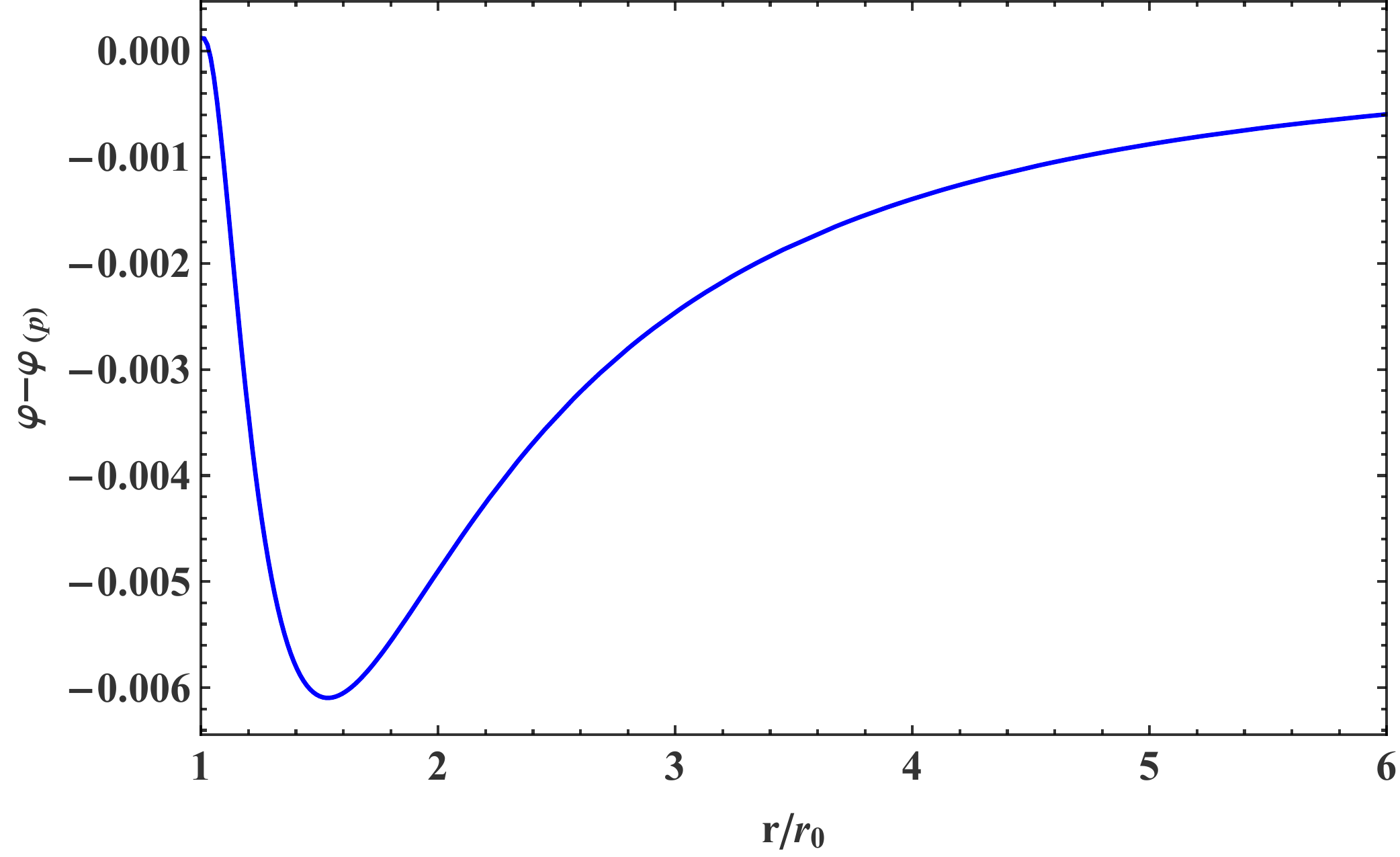}
\caption{For $p=0.5$ and at fourth order in the continued-fraction approximation, (left panel) the relative error in the $g_{rr}(r)$ metric function and (right panel) the error for the scalar field when $f(\vph)= \vph^2$.}
\label{fig:fDM Error for grr and phi}
\end{figure*}

In Fig.~\ref{fig:fDM Error for grr and phi} we plot the corresponding REs for both the $g_{rr}(r)$ metric function and the scalar field $\vph(r)$, both at the fourth order in the CFA. The expressions for the second-order analytical approximations for the metric functions and the scalar field can be found in Appendix A for all the GB couplings studied in this article\footnote{We only give the second-order expressions in the appendix for reasons of compactness but in the accompanying Mathematica\textregistered\, file one can obtain the analytical expressions up to fourth order. The file is available from \url{https://arxiv.org/src/1907.10112/anc/Approximation.nb}.}.

\subsubsection{The quartic GB-coupling functional}

The analytic approximations for the parameters of the CFA expansion in this case are

\beq
a_1= \frac{-\frac{115 p^3}{804}+\frac{13 p^2}{84}+\frac{p}{153}}{-\frac{85 p^3}{317}+p^2-\frac{168 p}{233}-\frac{13}{239}}\,,
\label{a1 even phi4}
\eeq

\beq
a_2= \frac{-\frac{207 p^3}{218}+p^2+\frac{85 p}{278}-\frac{1}{672}}{-\frac{57 p^2}{379}+\frac{43 p}{283}+\frac{1}{359}}\,,
\label{a2 even phi4}
\eeq

\beq
\eps= \frac{\frac{8 p^2}{247}+\frac{2 p}{331}}{\frac{111 p^2}{305}+p+\frac{28}{311}}\,,
\label{eps even phi4}
\eeq

\beq
b_1=\frac{\frac{53 p}{233}-\frac{26 p^2}{135}}{-\frac{73 p^2}{393}+p-\frac{282}{317}}\,,
\label{b1 even phi4}
\eeq

\beq
b_2=\frac{p^3-\frac{288 p^2}{397}-\frac{197 p}{308}-\frac{1}{142}}{\frac{39 p^2}{467}-\frac{32 p}{381}-\frac{1}{516}}\,,
\label{b2 even phi4}
\eeq

\beq
\vph_{\infty}=\frac{\frac{p^2}{33}+\frac{4 p}{27}}{p^2+\frac{193 p}{225}+\frac{1}{1136}}\,,
\label{phiInf even phi4}
\eeq

\beq
f_0 = \frac{-\frac{p^3}{140}+\frac{21 p^2}{136}+\frac{p}{61}}{p^2+\frac{29 p}{88}+\frac{1}{380}} \,,
\label{f0 even phi4}
\eeq

\beq
f_1 = \frac{\frac{99 p^2}{188}+\frac{11 p}{137}}{\frac{155 p^2}{182}+p+\frac{1}{62}} \,,
\label{f1 even phi4}
\eeq

\beq
f_2 = \frac{-\frac{32 p^4}{103}+p^3-\frac{39 p^2}{59}-\frac{92 p}{307}}{\frac{17 p^2}{97}-\frac{4 p}{19}-\frac{1}{1003}} \,,
\label{f2 even phi4}
\eeq

\subsection{The odd-polynomial coupling functional}
The odd-polynomial coupling functional is
\beq
f(\vph)= \vph^{2 n+1},\quad n \in \mathbb{N}\,.
\label{coupling: odd polynomial}
\eeq
For $\alp=1/4$ the dimensionless parameter has the following form:
\beq
p=6 (2 n+1)^2 \varphi_0 ^{4 n}\,,
\eeq
and the allowed values of $\vph_0$ are
\beq
|\vph_0 | \leqslant \left(6 (2 n+1)^2\right)^{-\frac{1}{4 n}}\,.
\eeq

For $n=1$, the approximate analytic expressions for the parameters are given below

\beq
a_1=\frac{\frac{11 p}{84}-\frac{79 p^2}{638}}{-\frac{124 p^2}{405}+p-\frac{233}{328}}\,,
\label{a1 odd}
\eeq

\beq
a_2=\frac{-\frac{179 p^3}{182}+p^2+\frac{97 p}{253}}{-\frac{41 p^2}{212}+\frac{41 p}{209}+\frac{1}{360}}\,,
\label{a2 odd}
\eeq

\beq
\eps = \frac{-\frac{p^3}{186}+\frac{7 p^2}{234}+\frac{3 p}{409}}{p+\frac{34}{271}}\,,
\label{eps odd}
\eeq

\beq
b_1 = \frac{\frac{33 p}{145}-\frac{58 p^2}{301}}{-\frac{77 p^2}{414}+p-\frac{217}{244}}\,,
\label{b1 odd}
\eeq

\beq
b_2 = \frac{p^3-\frac{119 p^2}{205}-\frac{186 p}{227}-\frac{1}{123}}{\frac{19 p^2}{181}-\frac{86 p}{811}-\frac{1}{512}} \,,
\label{b2 odd}
\eeq

\beq
\vph_{\infty} = \frac{\frac{p^3}{59}-\frac{10 p^2}{91}+\frac{5 p}{77}}{\frac{66 p^2}{83}+p+\frac{1}{1087}} \,,
\label{phiInf odd}
\eeq

\beq
f_0 = \frac{-\frac{p^3}{142}+\frac{7 p^2}{43}+\frac{2 p}{111}}{p^2+\frac{29 p}{75}+\frac{1}{312}} \,,
\label{f0 odd}
\eeq

\beq
f_1 = \frac{\frac{46 p^2}{91}+\frac{11 p}{138}}{\frac{101 p^2}{146}+p+\frac{2}{119}} \,,
\label{f1 odd}
\eeq

\beq
f_2 = \frac{-\frac{63 p^3}{80}+\frac{8 p^2}{117}+p+\frac{4}{143}}{\frac{8 p^3}{93}-\frac{51 p^2}{95}+\frac{59 p}{127}+\frac{2}{69}} \,,
\label{f2 odd}
\eeq

\subsection{The inverse-polynomial coupling functional}
The inverse-polynomial coupling functional is
\beq
f(\vph)= \vph^{-n},\quad n \in \mathbb{Z}^{+}\,,
\label{coupling: inverse polynomial}
\eeq
and the dimensionless parameter for $\alp=1/4$ turns out to be
\beq
p =6 n^2 \left( \vph_0 \right)^{-2 (n+1)} \,.
\eeq
The allowed range of values for $\vph_0$ in this case is
\beq
|\vph_0| \geqslant  \left( 6 n^2 \right)^{\frac{1}{2(n+1)}} \,.
\eeq
Once again, we fix $n=1$ in order to perform the analysis.

The approximate analytic expressions for the parameters in this case are given below

\beq
a_1 = \frac{\frac{13 p^3}{121}-\frac{4 p^2}{49}-\frac{16 p}{281}}{\frac{11 p^2}{355}-\frac{245 p}{263}+1} \,,
\label{a1 inv}
\eeq

\beq
a_2 = \frac{-\frac{583 p^3}{875}+p^2-\frac{31 p}{292}-\frac{13}{367}}{-\frac{57 p^2}{340}+\frac{49 p}{306}+\frac{2}{161}} \,,
\label{a2 inv}
\eeq

\beq
\eps = \frac{-\frac{2 p^3}{317}+\frac{25 p^2}{341}+\frac{8 p}{147}}{p+\frac{167}{308}} \,,
\label{eps inv}
\eeq

\beq
b_1 = \frac{\frac{103 p}{477}-\frac{75 p^2}{406}}{-\frac{12 p^2}{59}+p-\frac{165}{191}} \,,
\label{b1 inv}
\eeq

\beq
b_2 = \frac{-\frac{300 p^2}{307}+\frac{585 p}{1756}+1}{\frac{25 p^2}{182}-\frac{133 p}{257}+\frac{259}{671}} \,,
\label{b2 inv}
\eeq

\beq
\vph_{\infty} = \frac{-\frac{25 p^3}{84}+p^2+\frac{31 p}{113}+\frac{1}{181}}{-\frac{11 p^3}{69}+\frac{43 p^2}{84}+\frac{7 p}{83}+\frac{1}{1297}} \,,
\label{phiInf inv}
\eeq

\beq
f_0 = \frac{-\frac{p^3}{71}-\frac{28 p^2}{71}-\frac{5 p}{141}}{p^2+\frac{27 p}{58}+\frac{1}{198}} \,,
\label{f0 inv}
\eeq

\beq
f_1 = \frac{-\frac{10 p^3}{143}-\frac{27 p^2}{190}-\frac{p}{194}}{p^2+\frac{2 p}{15}+\frac{1}{1496}} \,,
\label{f1 inv}
\eeq

\beq
f_2 = \frac{\frac{3 p^3}{16}-\frac{7 p^2}{81}-\frac{7 p}{71}-\frac{1}{46}}{-\frac{41 p^3}{97}+p^2-\frac{121 p}{230}-\frac{1}{17}} \,,
\label{f2 inv}
\eeq

\subsection{The logarithmic coupling functional}
Finally we turn to the logarithmic coupling functional
\beq
f(\vph)= \ln{(\vph)} \,,
\label{coupling: log}
\eeq
the dimensionless parameter for $\alp=1/4$ is
\beq
p = \frac{6}{\vph_0^2}\,,
\eeq
and the allowed values of $\vph_0$ are
\beq
|\vph_0| \geqslant \sqrt{6} \,.
\eeq

The approximate analytic expressions for the parameters in this case are given below

\beq
a_1 = \frac{\frac{11 p^3}{761}-\frac{113 p^2}{2844}+\frac{18 p}{709}}{-\frac{605 p^2}{1314}+p-\frac{425}{787}} \,,
\label{a1 log}
\eeq

\beq
a_2 = \frac{-\frac{586 p^3}{813}+\frac{367 p^2}{666}+p-\frac{384}{611}}{\frac{237}{1072}-\frac{255 p}{1189}}\,,
\label{a2 log}
\eeq

\beq
\eps = \frac{\frac{105 p}{1018}-\frac{44 p^2}{1323}}{1-\frac{349 p}{657}} \,,
\label{eps  log}
\eeq

\beq
b_1 = \frac{\frac{85 p}{387}-\frac{57 p^2}{308}}{-\frac{29 p^2}{152}+p-\frac{222}{251}} \,,
\label{b1 log}
\eeq

\beq
b_2 = \frac{-\frac{66 p^2}{197}+p-\frac{262}{383}}{-\frac{73 p^2}{318}+\frac{149 p}{295}-\frac{69}{250}} \,,
\label{b2 log}
\eeq

\beq
\vph_{\infty} = \frac{-\frac{277 p^3}{405}+\frac{159 p^2}{218}+p+\frac{10}{253}}{-\frac{37 p^3}{458}+\frac{167 p^2}{246}+\frac{73 p}{608}+\frac{1}{1310}} \,,
\label{phiInf log}
\eeq

\beq
f_0 = \frac{\frac{140 p^3}{449}+\frac{58 p^2}{271}+\frac{p}{232}}{p^2+\frac{25 p}{214}+\frac{1}{3196}}\,,
\label{f0 log}
\eeq

\beq
f_1 = \frac{-\frac{83 p^3}{188}+\frac{160 p^2}{239}+\frac{28 p}{285}}{-\frac{269 p^2}{372}+p+\frac{3}{148}} \,,
\label{f1 log}
\eeq

\beq
f_2 = \frac{\frac{81 p^3}{203}+\frac{37 p^2}{378}-\frac{350 p}{571}-\frac{29}{802}}{-\frac{139 p^3}{605}+p^2-\frac{173 p}{233}-\frac{31}{384}} \,,
\label{f2 log}
\eeq

At this point, one must mention an important phenomenon, the \emph{eikonal instability}, which takes place when the Gauss-Bonnet term is turned on.
Once the Gauss-Bonnet coupling constant is not small enough, the black-hole solution suffers from a dynamical instability: if linearly perturbed, the perturbation grows unboundedly. The linear instability breaks down in the regime of small perturbations, indicating that the black hole cannot exist in this range of parameters.

The instability brought by the Gauss-Bonnet term is of special kind: it develops at high multipole numbers, so that the summation over the multipole numbers cannot be valid anymore. This kind of instability was first observed for the higher-dimensional Einstein-Gauss-Bonnet black holes \cite{Dotti:2005sq} and later observed for a number of other cases, including black branes \cite{Takahashi:2011du}, asymptotically de Sitter and anti-de Sitter black holes \cite{Cuyubamba:2016cug,Konoplya:2017ymp,Konoplya:2017zwo}, black holes and branes in theories with higher than the second order in curvature corrections \cite{Takahashi:2010gz,Takahashi:2011qda,Grozdanov:2016fkt,Konoplya:2017lhs}. In some cases, the instability occurs not only for the gravitational perturbations, but also for the test scalar field \cite{Gonzalez:2017gwa}.

As the eikonal instability is a very wide phenomenon which, it seems, does not depend on a particular form of the higher-curvature correction, we believe that it must be present also for the Einstein-scalar-Gauss-Bonnet theory at least once the scalar coupling is strong enough. Therefore, the regime of near extremal $p$, corresponding to the maximal coupling, most probably does not represent any realistic stable black hole. Exactly in this regime our continued-fraction expansion converges slowly. On the contrary, in the regime where one can expect stable configuration the second-order expansion is sufficient to constrain the relative error by a fraction of one percent. In other words, our analytical approximation is very accurate already at the second order, once one is limited by stable configurations.

\section{Black-hole shadows and accuracy of the analytical approximation}\label{sec:shadows}

In the previous sections we have obtained approximate analytical expressions for the metric functions and the scalar field up to fourth order in the CFA. In all cases, we have found excellent agreement between the numerical and analytical solutions by computing the RE. Still, the metric itself is not gauge invariant and comparison of various metric functions does not allow us to determine the accuracy of the analytical approximation. For the latter one needs to consider some gauge-invariant, observable quantity.

Recently black-hole shadows have been intensively studied for various theories of gravity and astrophysical environment (an extensive, but not exhaustive list of works can be found in \cite{Contreras:2019cmf,Tsukamoto:2017fxq,Mizuno:2018lxz,Zhu:2019ura,Dokuchaev:2018kzk,Xu:2018mkl,Ovgun:2018tua,Huang:2018rfn,Hou:2018avu,Wei:2018xks,Davelaar:2018dfp,Dokuchaev:2018ibr,Mishra:2019trb,Abdikamalov:2019ztb,Held:2019xde,Shaikh:2019fpu,Wang:2019tjc,Ali:2019khp,Long:2019nox,Ovgun:2019jdo,Contreras:2019cmf,Wang:2017hjl} and references therein). In this section we perform the computation of the shadows cast by the EsGB black holes numerically. For different orders in the continued-fraction approximation we compute the shadows and compare them with the numerical ones. This way we have a gauge-invariant measure of the accuracy of our approximation.

\begin{figure*}
\includegraphics[angle=0.0,width=0.5\linewidth]{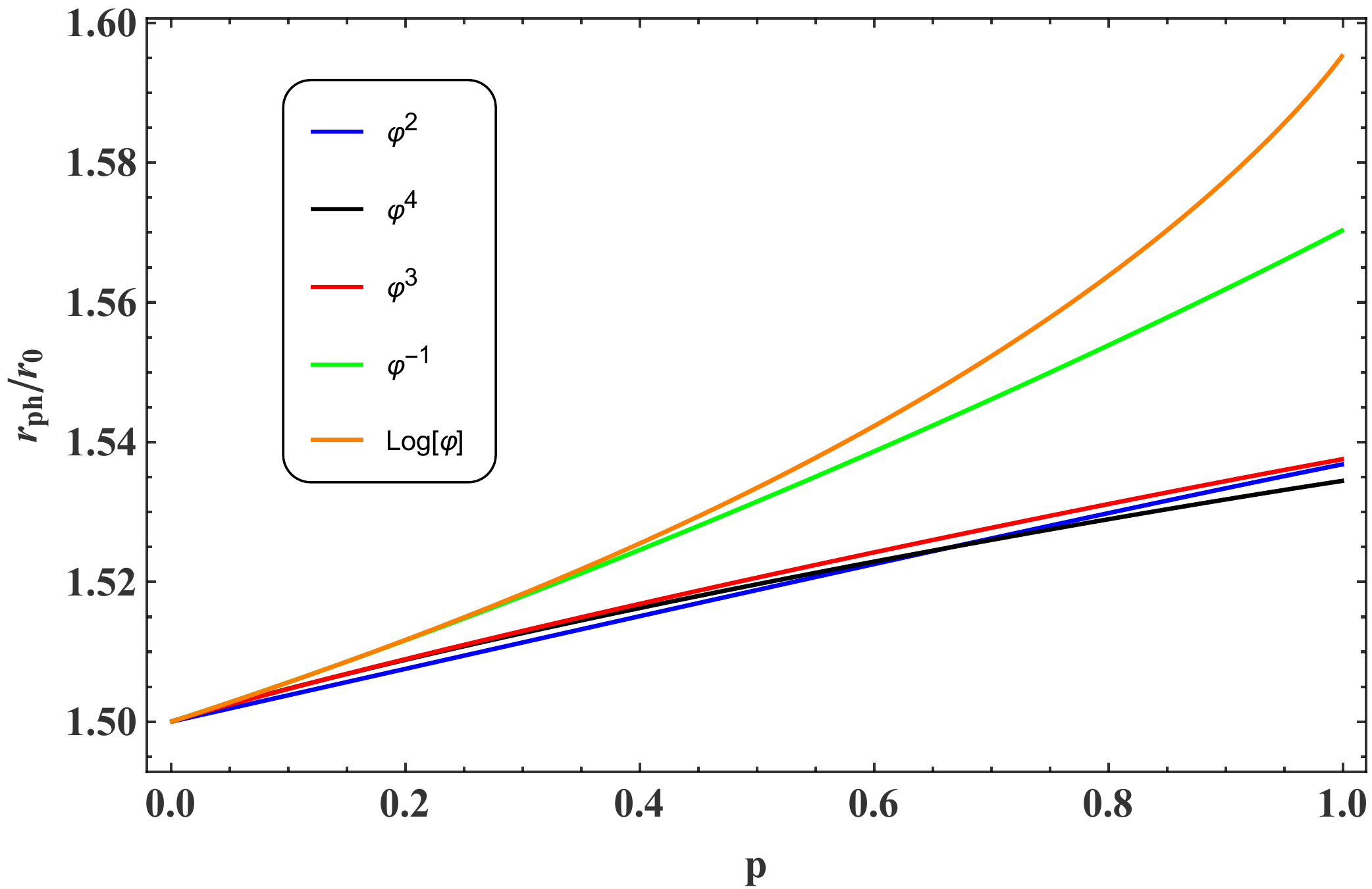}\includegraphics[angle=0.0,width=0.5\linewidth]{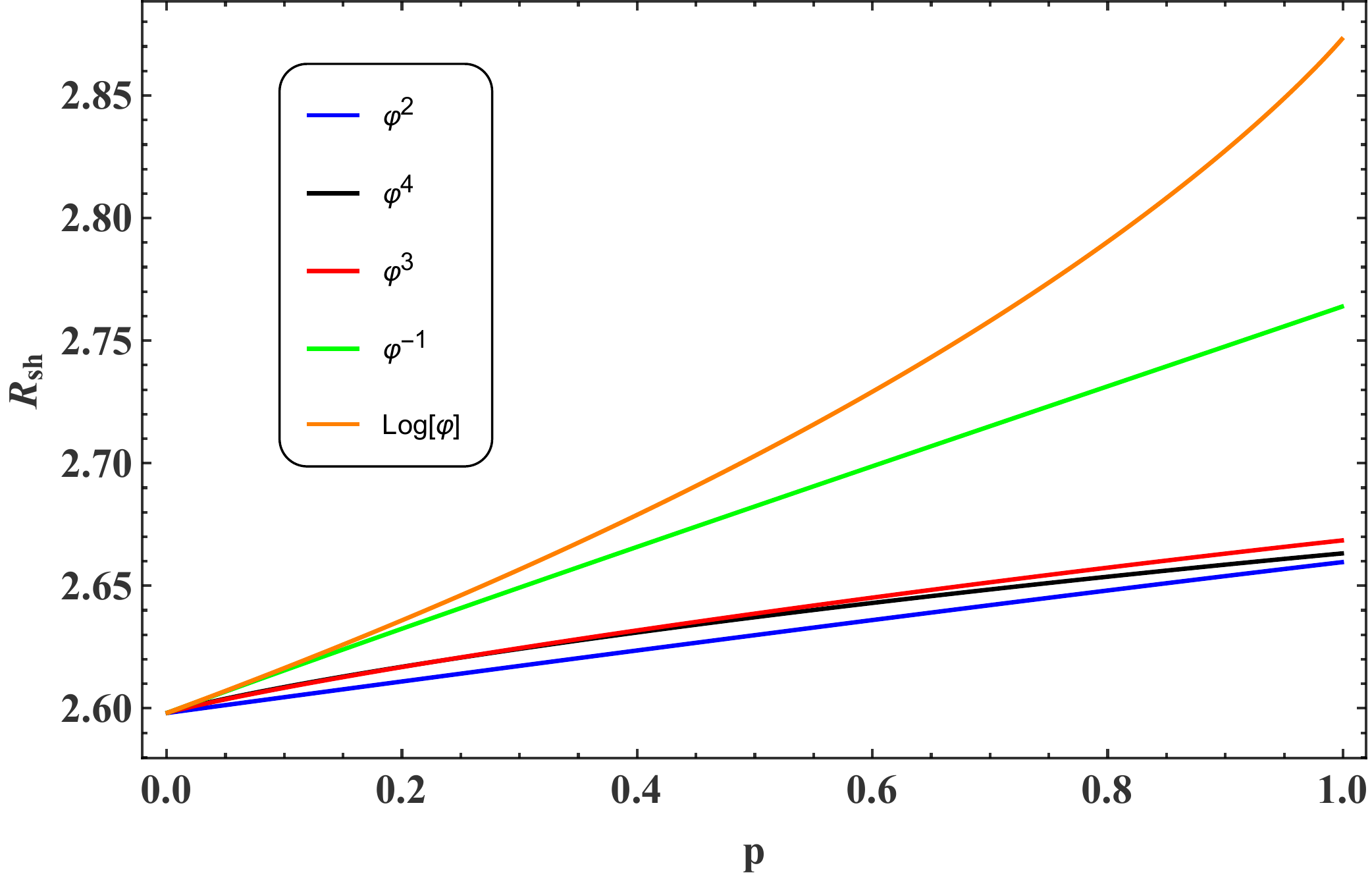}
\caption{The numerical values for the radius of the photon sphere (left panel) and the black-hole shadow (right panel) for each value of $p$.}
\label{fig: rph and shadows all cases}
\end{figure*}

The radius of the photon sphere $r_{ph}$ of a black hole in the coordinate system of \eqref{metric ansatz} is determined by means of the following function (see, for example, \cite{Perlick2015,Konoplya2019} and references therein):
\beq
h^2(r) \equiv \frac{r^2}{g_{tt}(r)} \,,
\label{h2 definition}
\eeq
as the solution to the equation
\beq
\frac{d}{dr} h^2 (r)=0\,.
\eeq
Then, the radius of the black-hole shadow $R_{sh}$ as seen by a distant static observer located at $r_O$ is
\beq
R_{sh} = \frac{h(r_{ph})r_O}{h(r_O)} = \frac{r_{ph}\sqrt{g_{tt}(r_O)}}{\sqrt{g_{tt}(r_{ph})}} \approx  \frac{r_{ph}}{\sqrt{g_{tt}(r_{ph})}}\,,
\label{shadow def}
\eeq
where in the last equation we have assumed that the observer is located sufficiently far away from the black hole so that she/he is deep in the asymptotically flat regime, i.e. $g_{tt}(r_O) \approx 1$.

In the case of the Schwarzschild black hole it is known that $r_{ph}= 1.5 \, r_0$ and so according to \eqref{shadow def} the shadow is $R_{sh} \approx 2.59808 \, r_0$. For the EsGB black holes, the deviations from these two limiting values are expected to increase with the parameter $p$ as we move further and further away from the Schwarzschild limit $(p=0)$. This is indeed the case as the plots for the numerical values of $r_{ph}$ and $R_{sh}$ reveal in Fig.~\ref{fig: rph and shadows all cases}. We point out that although $r_{ph}$ is a nonobservable auxiliary quantity, which is not gauge invariant, it is very useful in many applications beyond the computation of black-hole shadows. To this end, its profile with $p$ as depicted in Fig.~\ref{fig: rph and shadows all cases} provides useful information. Also, in Appendix C, the interested reader can find approximate analytical expressions for these two quantities.

Having obtained the accurate solutions for the shadows numerically we can now compare how each order in the CFA stands against the numerical solutions. By terminating the series of the expansion of $\tilde{A}(x)$ \eqref{contfrac} each time at $a_1$, $a_2$, $a_3$, $a_4$, and $a_5$ we obtain the first-, second-, third-, fourth-, and fifth-order analytical approximation for the $g_{tt}(r)$ metric function, respectively.

\begin{table}
\begin{tabular}{llllll}
\hline
             &$f=\phi^2$   & $f=\phi^3$  & $f=\phi^4$  & $f=\ln(\phi)$ & $f=\phi^{-1}$ \\
             &$p=0.5$      & $p=0.3$     & $p=0.4$     & $p=0.5$       & $p=0.9$       \\
\hline
$R_{sh}/r_0$ & $2.62984 $  & $2.62456 $  & $2.63095$   & $2.70299$     & $2.74766 $    \\
$RE_1$       & $1.3759\%$  & $0.5597\%$  & $0.8669\%$  & $0.1694\%$    & $1.5947\%$    \\
$RE_2$       & $0.0782\%$  & $0.0111\%$  & $0.0138\%$  & $0.0198\%$    & $0.7590\%$    \\
$RE_3$       & $0.3380\%$  & $0.0625\%$  & $0.1307\%$  & $0.4864\%$    & $0.6678\%$    \\
$RE_4$       & $0.1224\%$  & $0.0214\%$  & $0.0188\%$  & $0.0495\%$    & $0.4288\%$    \\
$RE_5$       & $0.0413\%$  & $0.0015\%$  & $0.0104\%$  & $0.0105\%$    & $0.3065\%$    \\
\hline
\end{tabular}
\caption{The accurate value of the black-hole shadow radius and the relative error for the approximations of the first five orders.}\label{tabl:convergence}
\end{table}

The absolute RE of the analytical approximation of the black-hole shadow from the numerical solution $R_{sh}$ for different couplings and values of the dimensionless parameter $p$ are given in Table~\ref{tabl:convergence}. Starting from the fourth-order approximation the relative error quickly decreases for small $p$. For larger values of $p$ the convergence becomes slower, so that for the near-extremal black holes we need more orders to achieve a reasonable approximation. However, for the nonextremal black holes, the second- and fourth-order approximations give RE of fractions of a percent. Notice that the third-order approximation usually leads to a slightly worse accuracy than those of the second and fourth order.

The analytic expressions for the metric functions in the second and fourth order in CFA deviate from the numerical ones by less than $1\%$ for almost the entirety of the GB couplings that we have studied in this work. Only for the inverse polynomial case, the deviation is slightly larger, but still smaller than $ 1.4\%$. It is noteworthy that these maximal values for the RE actually emerge in the large-$p$ regime where the black holes are presumably unstable. Thus for viable black-hole solutions the RE is quite small.

\section{On the approximation for different values of the coupling constant}

In the analysis performed in this article we considered $\alp=1/4$ in order to perform the fitting and obtain the approximate analytical expressions for the metric functions and the scalar field. In principle with the help of the analytical expressions obtained here one is able to consider a range of values for the GB constant $\alpha$.

\begin{figure*}
\includegraphics[width=0.5\linewidth]{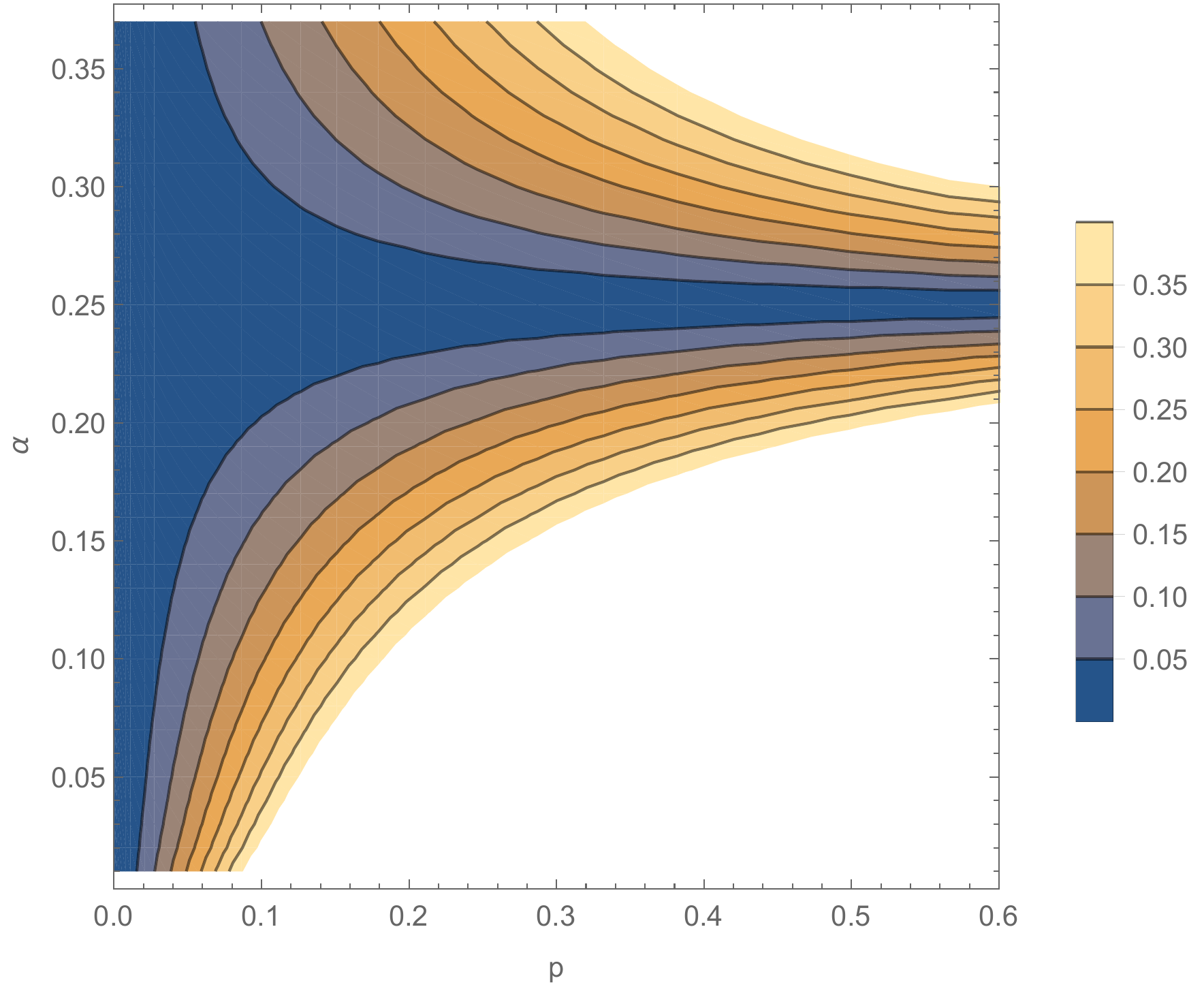}\includegraphics[width=0.5\linewidth]{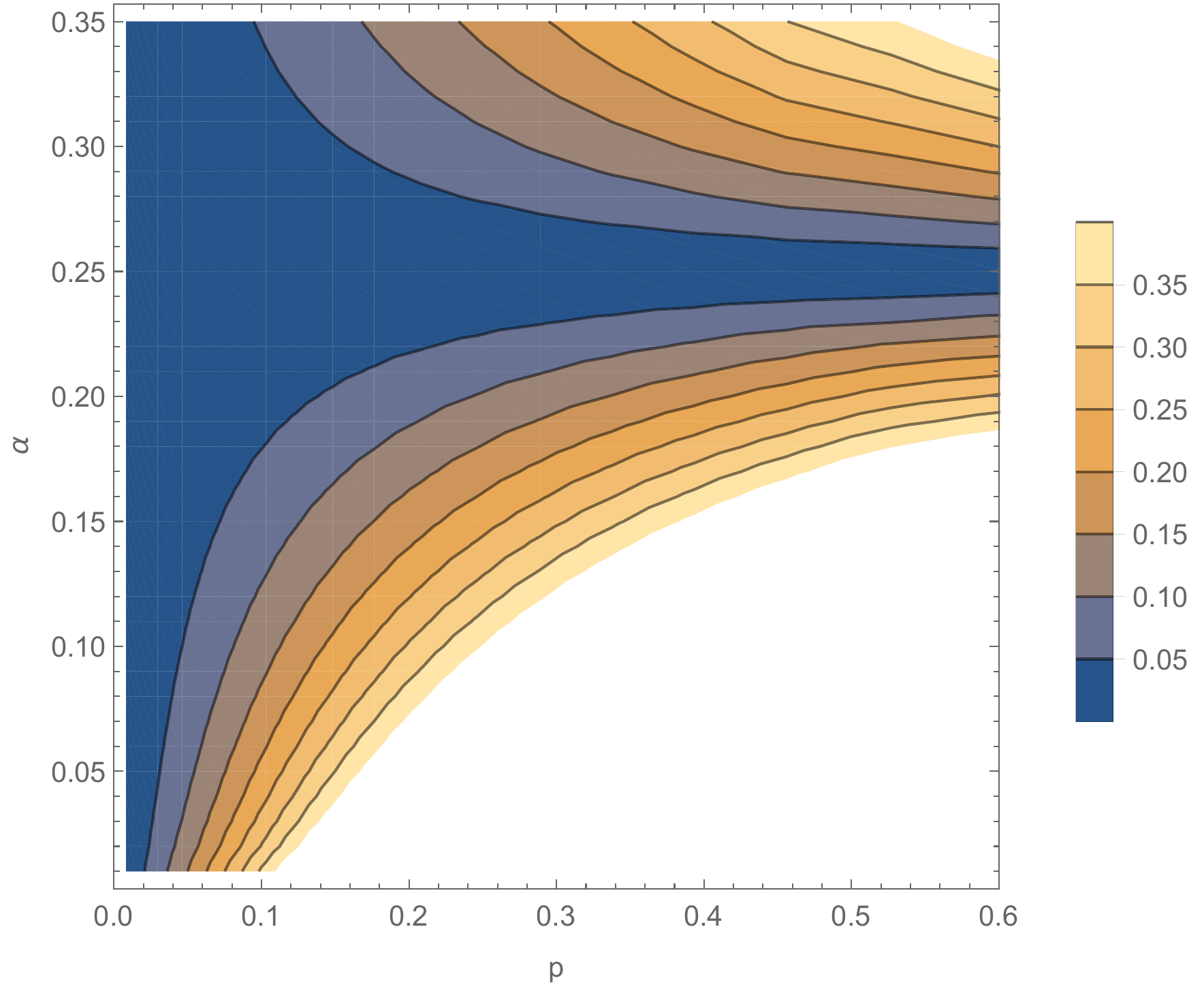}
\includegraphics[width=0.5\linewidth]{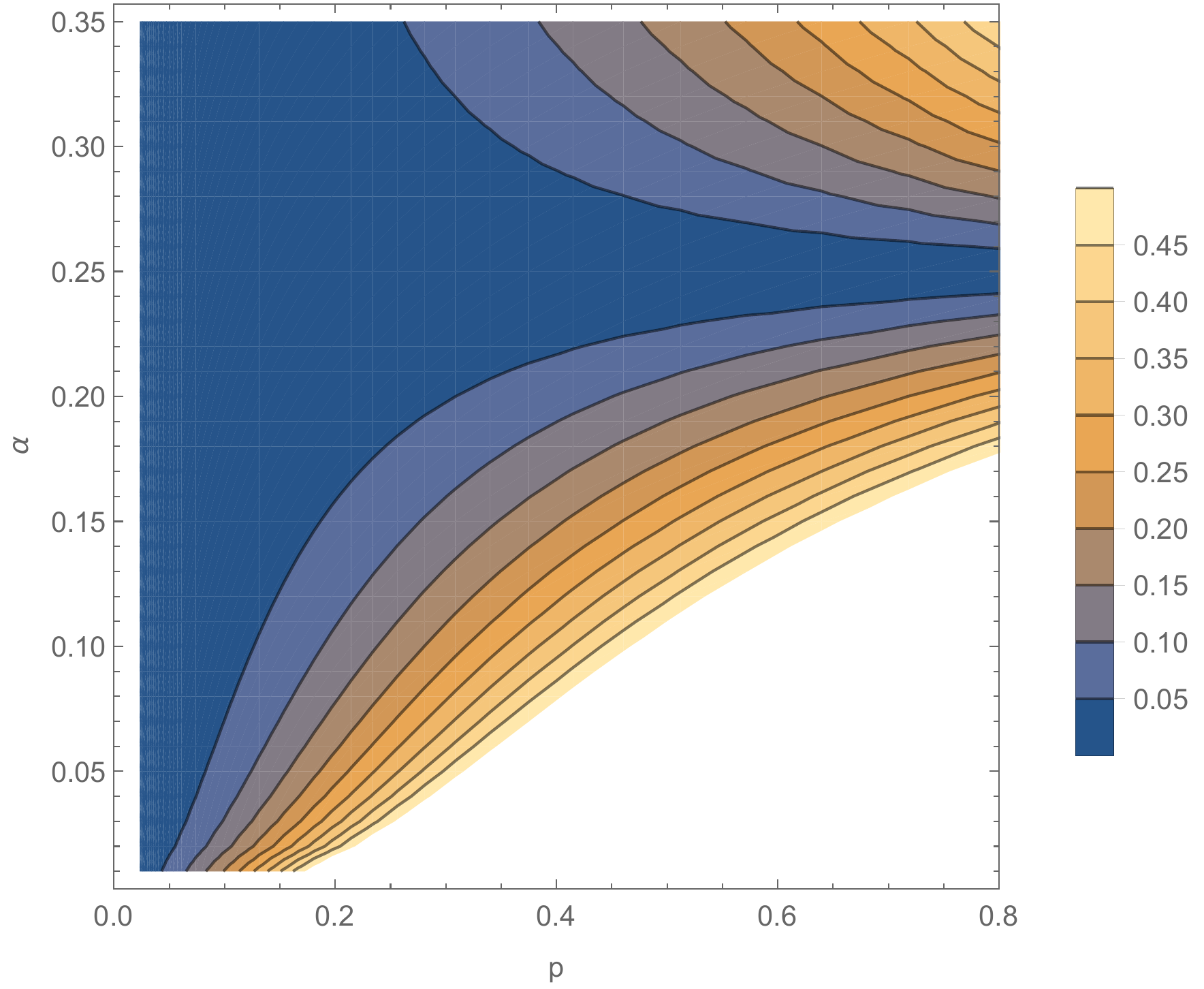}\includegraphics[width=0.5\linewidth]{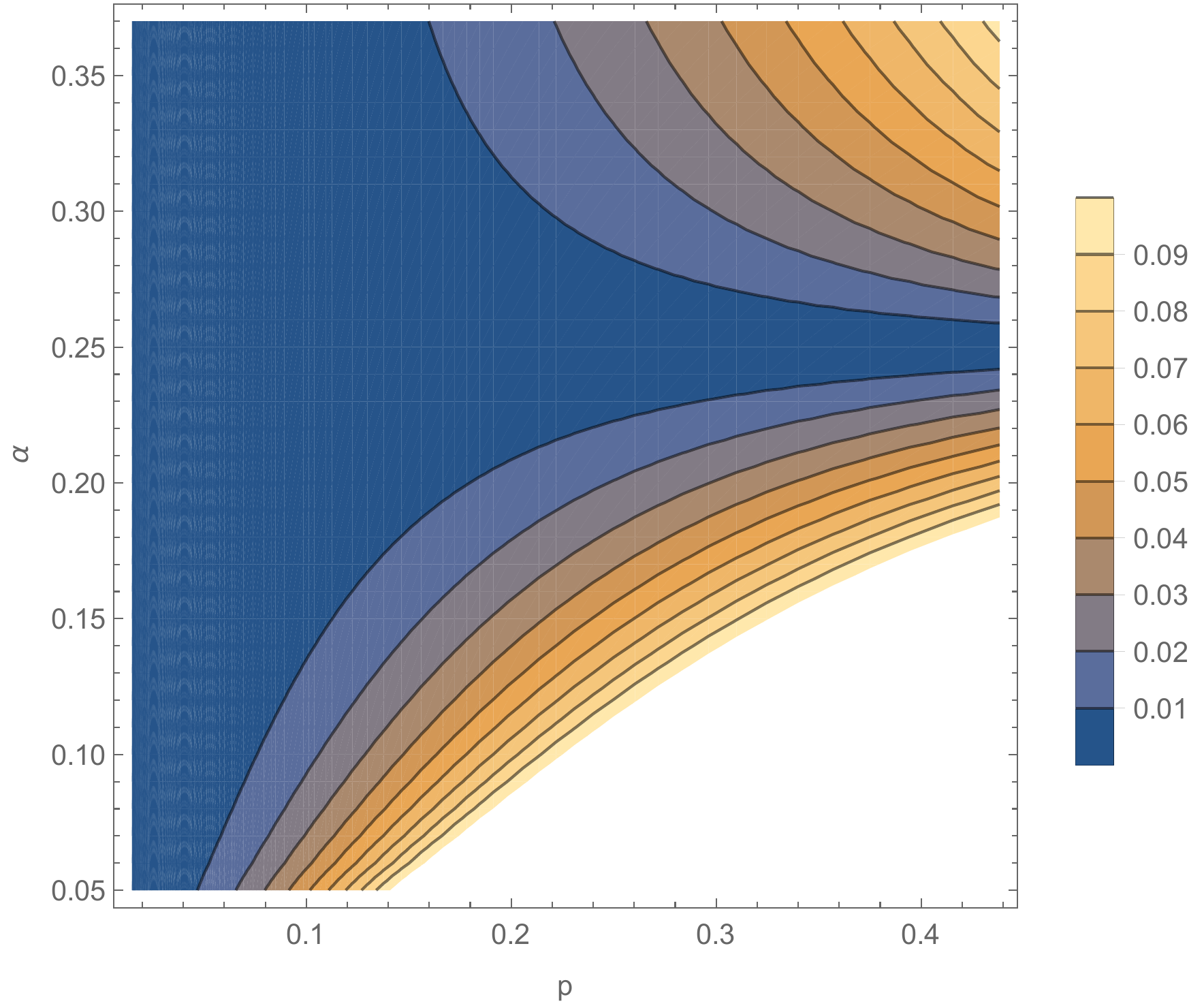}
\caption{Maximum relative difference (in percents) for the function $g_{tt}(r)$ between the numerical solution for given values of $\alp$ and $p$ and the numerical solution obtained for $\alp=1/4$ and the same value of $p$, found for the following coupling functionals: $f(\phi) = \vph^3$ (top left), $f(\phi) = \vph^4$ (top right), $f(\phi) = \vph^{-1}$ (bottom left), and $f(\phi) = \ln{(\vph)}$ (bottom right).} \label{fig:MRE_metric}
\end{figure*}

By considering other values of the coupling constant $\alp$ we have observed that for small $p$ the variation of the metric function is negligible. From Fig.~\ref{fig:MRE_metric} we see that the relative difference between the accurate metric function and the function obtained by fixing $\alp=1/4$ is as small as fraction of a percent when $p\lesssim0.2$.

We are now in a position to find the appropriate approximation for the scalar field $\vph(r)$ when $\alp\neq1/4$. First, we can introduce
\beq
\bar{\vph}(r)=C_{\alp}\vph(r),
\eeq
where $C_{\alp}$ is a constant, such that
\beq\label{Cscale}
\frac{1}{4}\frac{f'(\bar{\vph})}{\bar{\vph}}=\alp\frac{f'(\vph)}{\vph}.
\eeq
Assuming, that the metric depends on the parameter $p$ only, we notice that $\bar{\vph}(r)$ satisfies \eqref{phi-eq_0} for $\alp=1/4$, i.e. can be well approximated by \eqref{Fx} with some effective value of the parameter $p$,
\beq
p = 6 f'(\bar{\vph}_0)^2=96\alp^2C_{\alp}^2f'(\vph_0)^2=C_{\alp}^2 p_{\alp}.
\eeq
Here $p_{\alp}$ is the actual value of the parameter given by \eqref{dimensionless param}, which should be used in the expression for the metric functions.

For any of the polynomial couplings $f(\vph)=\vph^r$ Eq.~\eqref{Cscale} yields
\beq\label{Cscalesol}
C_{\alp}=\left(4\alpha\right)^{\frac{1}{r-2}},
\eeq
and, similarly, for $f(\vph)=\ln(\vph)$, we take $r=0$ in \eqref{Cscalesol}.

\begin{figure*}
\includegraphics[width=0.5\linewidth]{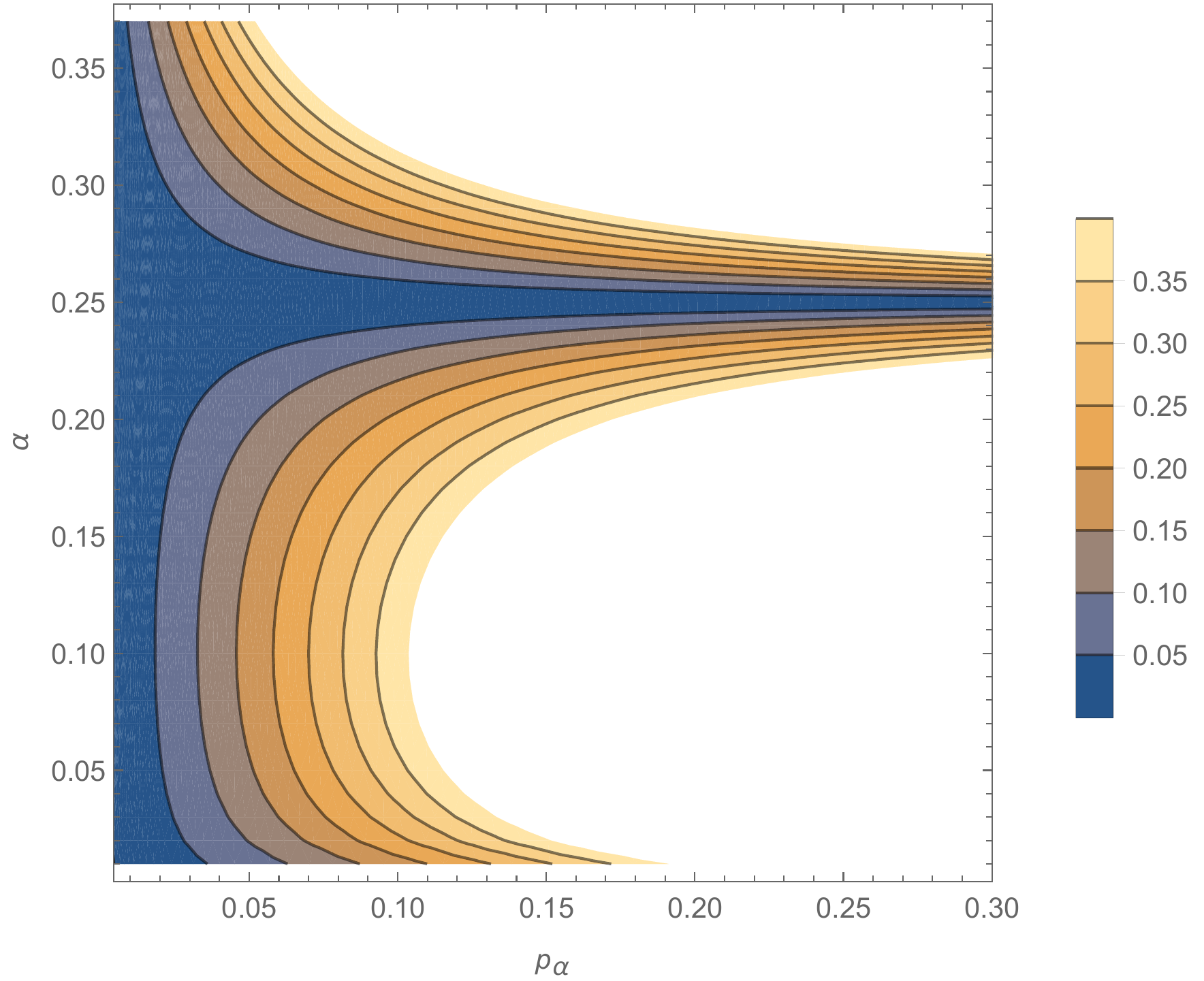}\includegraphics[width=0.5\linewidth]{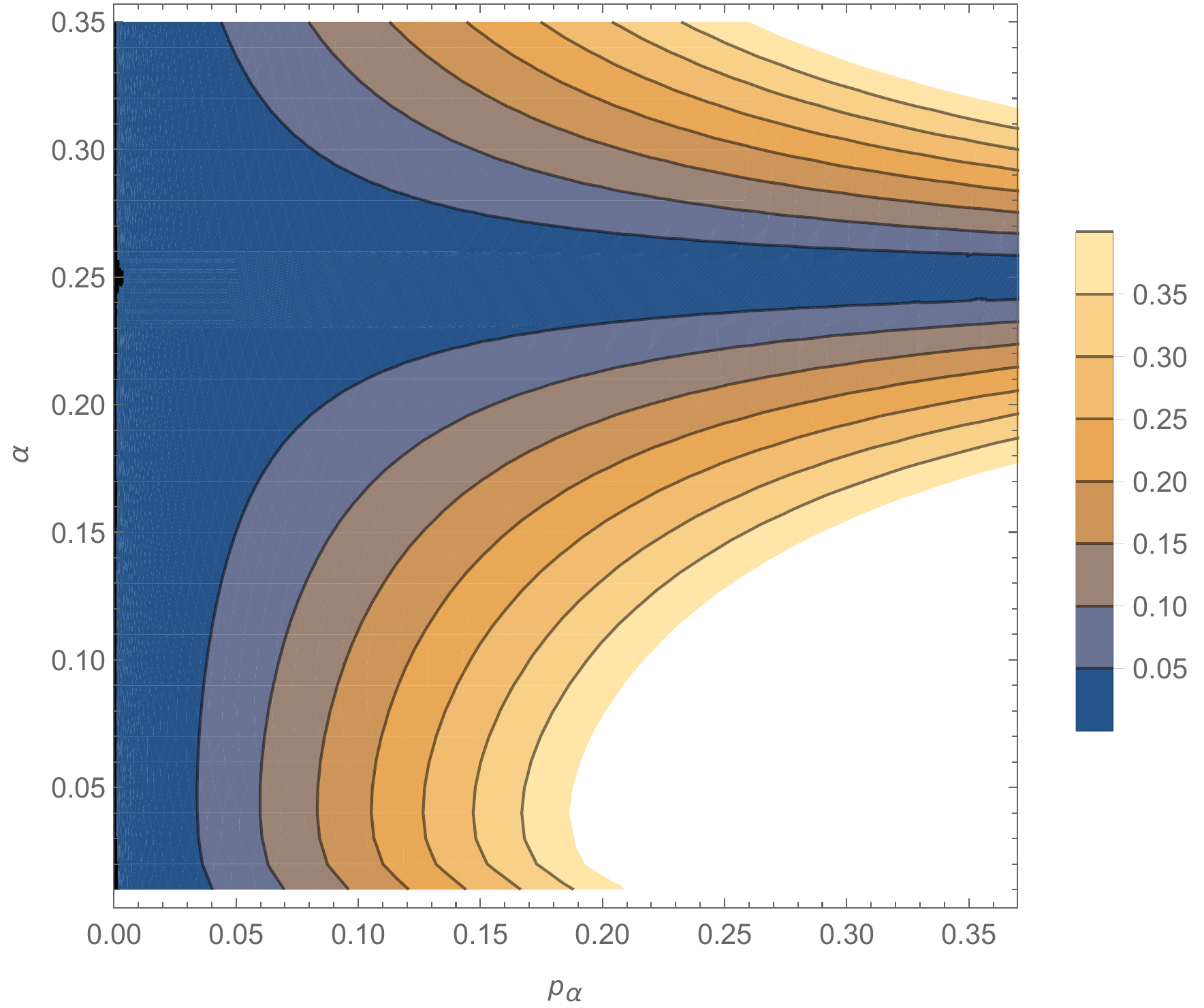}
\includegraphics[width=0.5\linewidth]{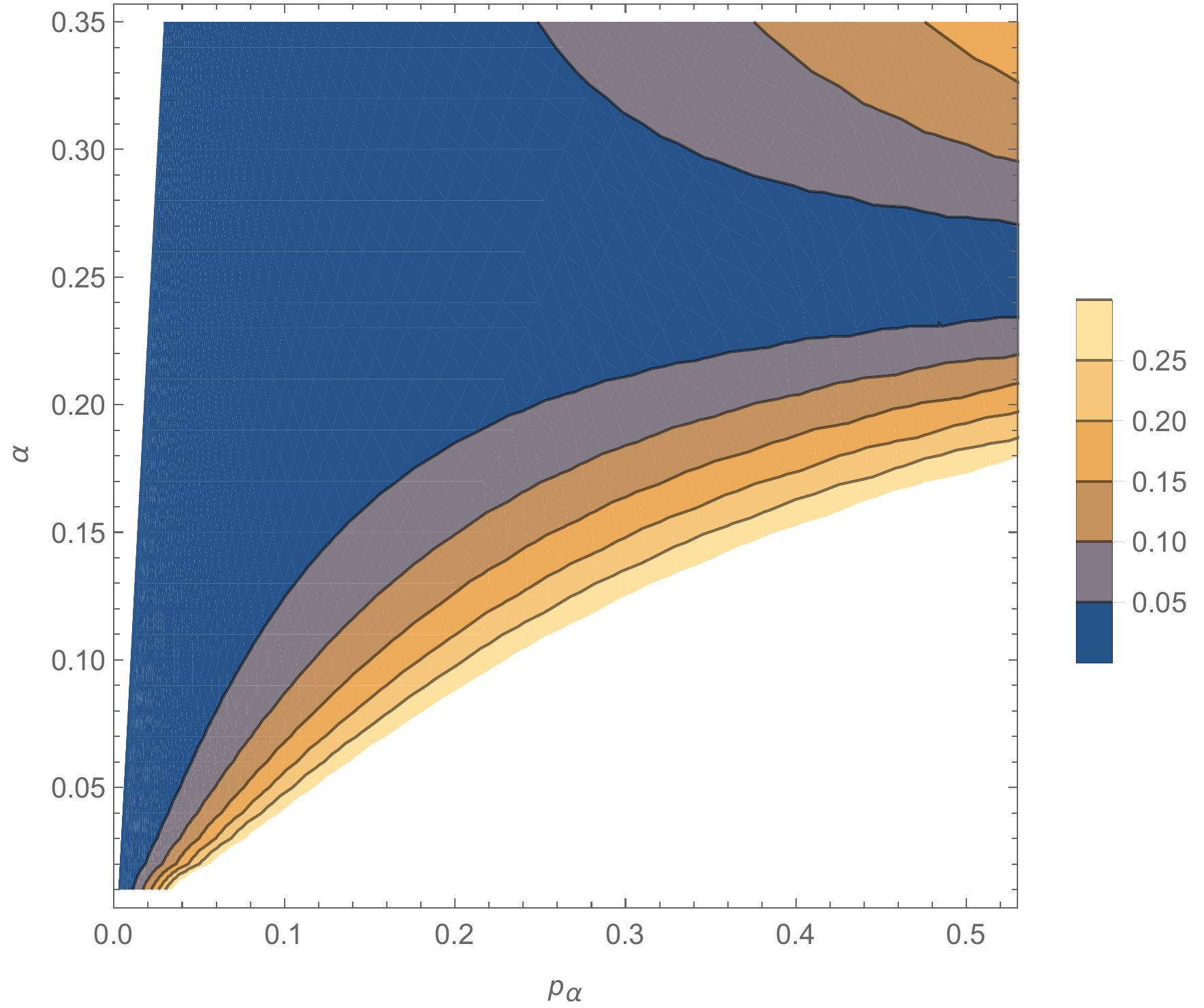}\includegraphics[width=0.5\linewidth]{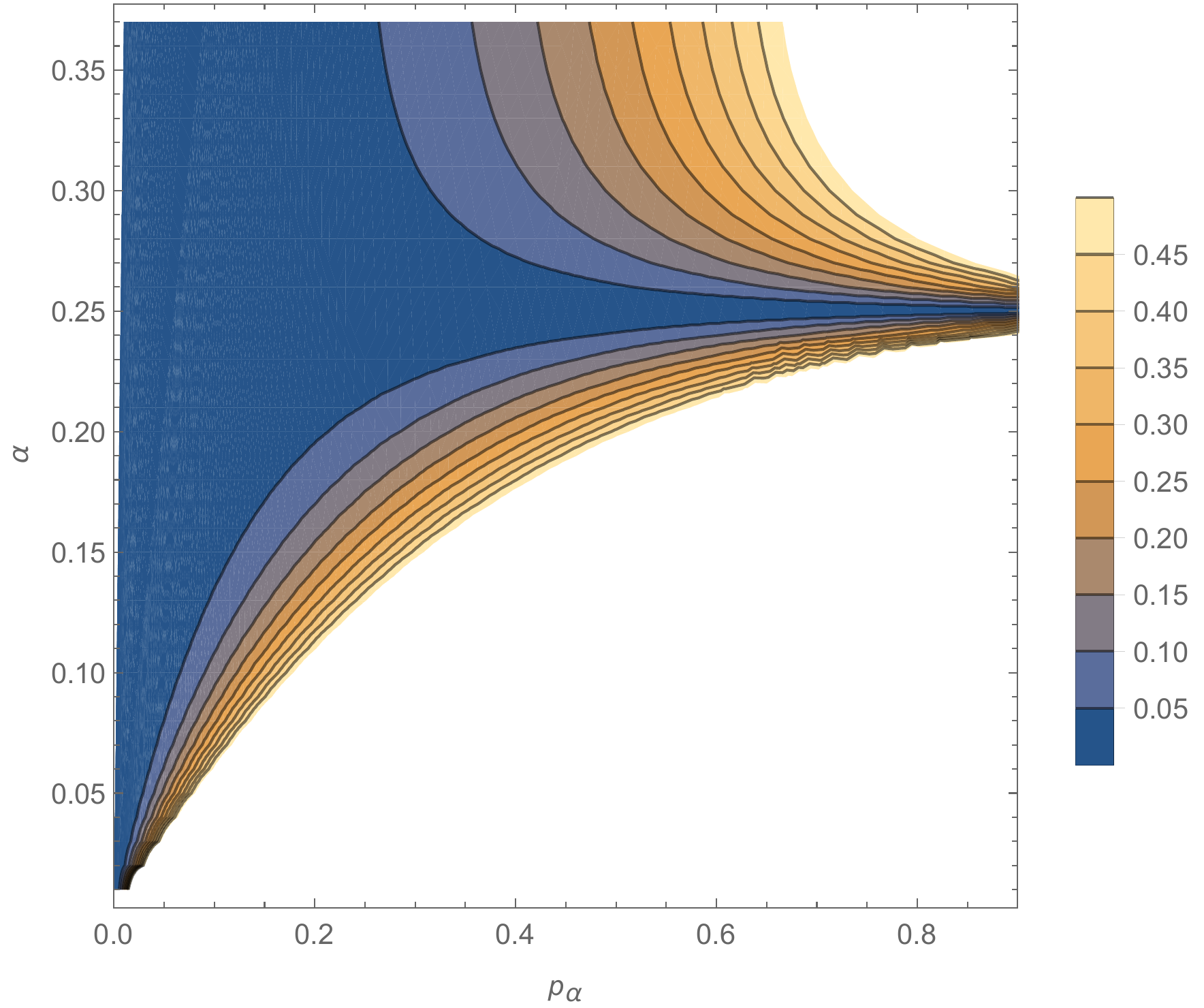}
\caption{Maximum relative difference (in percents) between $C_{\alp}\vph(r)$ for given values of $\alp$ and $p_{\alp}$ and $\bar{\vph}(r)$ obtained for $\alp=1/4$ and the corresponding effective value of $p$, found numerically for the following coupling functionals: $f(\phi) = \vph^3$ (top left), $f(\phi) = \vph^4$ (top right), $f(\phi) = \vph^{-1}$ (bottom left), and $f(\phi) = \ln{(\vph)}$ (bottom right).} \label{fig:MRE_phi}
\end{figure*}

On Fig.~\ref{fig:MRE_phi} we show the relative error of the above approximation for the scalar field. Namely, we compare $C_{\alp}\vph(r)$ for various values of $\alp$ and $p_{\alp}$ and the function $\bar{\vph}(r)$ for $\alp=1/4$ and the corresponding effective value of $p$,
$$p=C_{\alp}^2 p_{\alp}=\left(4\alpha\right)^{\frac{2}{r-2}}p_{\alp}.$$
We conclude that the scalar field can be approximated as
\beq
  \vph(r)\approx\frac{1}{C_{\alp}}\bar{\vph}(r),
\eeq
when $p_{\alp}$ is sufficiently small.

The only exception is the coupling functional $f(\phi) = \vph^2$, for which (\ref{Cscale}) cannot be satisfied. Nevertheless, the obtained approximation for the metric functions can still be used in this case.

\section{Conclusions}

In the context of Einstein-scalar-Gauss-Bonnet gravity, a plethora of black hole solutions with nontrivial scalar hair emerge for different coupling functionals to the Gauss-Bonnet term \cite{Antoniou2018a}. This has been recently demonstrated in \cite{Antoniou2018} where numerical solutions to the field equations have been obtained for four different GB couplings (even-, odd-, inverse-polynomial and logarithmic). In this work, we employed the powerful method of the continued-fraction approximation \cite{Rezzolla2014} in order to obtain analytic expressions for the metric functions and the scalar field for the aforementioned GB couplings.

For each coupling functional we parametrized the family of black-hole solutions that emerge in terms of a dimensionless compact parameter $p$ that ranges from $0$ (Schwarzschild limit) to $1$. The analytical representation is based on the continued-fraction expansion which converges quickly for all values of $p$ except the regime of near extremal coupling, when $p$ is close to unity. It is known that in this regime, Gauss-Bonnet black holes (as well as all the other known higher-curvature corrected black holes and branes whose gravitational perturbations were investigated) are unstable and, therefore, cannot exist. Although the (in)stability for the above considered couplings of the scalar field have not been studied in the literature so far, we assume that at least in the regime of the strong scalar field, the instability should remain. It would be interesting to check this supposition on the instability of EsGB black holes in the future and the obtained here analytical approximations for the black-hole metric and scalar field makes further investigation of stability easier.

We performed the computation up to the fourth order in the continued-fraction expansion and we have found that the deviation of the analytic expressions from the accurate numerical ones is at most of the order of $\mathcal{O}(1)\, \%$ for black-hole configurations which are expected to be gravitationally stable. This observation alone is not sufficient to guaranty the high accuracy of the approximation since the metric coefficients are not gauge-invariant quantities.

To this end, in order to make a concrete and gauge-invariant statement about the accuracy of the approximation we turned to the black-hole shadows cast by the EsGB black holes. We computed the shadows for five GB couplings numerically and compared them against the approximate results obtained via the analytical approximation to second, third and fourth order. We found that already in the second order, the largest relative error for the analytical approximations emerges in the maximal coupling limit $p \rightarrow 1$ and is less than $1\, \%$.

We noticed that all the considered coupling functionals lead to an increase of the radius of the black-hole shadow with respect to the Schwarzschild value $R_{sh}\approx2.59808 r_0$. In addition, we have obtained analytical expressions for the photon sphere which increases for all the couplings as well. The analytical representation obtained here for the black-hole metrics and scalar fields in the Einstein-scalar-Gauss-Bonnet theory allows one to explore various analytical, semianalytical, and numerical tools in order to  study various effects in the background of these solutions, such as accretion of matter, quasinormal modes, scattering, Hawking radiation and others.

\begin{acknowledgments}
We thank G. Antoniou, A. Bakopoulos, and P. Kanti for sharing their numerical code with us. A.~Z. was supported by Conselho Nacional de Desenvolvimento Científico e Tecnológico (CNPq). The authors acknowledge the support of the grant 19-03950S of Czech Science Foundation (GAČR). This publication has been prepared with partial support of the ``RUDN University Program 5-100''.
\end{acknowledgments}

\appendix
\section{Analytical expressions for the metric functions and the scalar field to second order in the CFA}
\label{App A}
In this appendix we give the explicit expressions for the analytical approximations for the metric functions and the scalar field in second order in the CFA.

These functions are rational functions of $r$, which we give in the following form:
\begin{eqnarray}
g_{tt}(r)&\approx & \mathcal{N}^{(1)}/\mathcal{D}^{(1)} \left(1-\frac{r_0}{r} \right) \,,
\\
\sqrt{g_{tt}(r) g_{rr}(r)}& \approx & \mathcal{N}^{(2)}/\mathcal{D}^{(2)}\,,
\\
e^{\vph (r)-\vph_\infty} &\approx & \mathcal{N}^{(3)}/\mathcal{D}^{(3)}\,,
\end{eqnarray}

where the numerators, $\mathcal{N}^{(1)},\mathcal{N}^{(2)},\mathcal{N}^{(3)},$ and the denominators, $\mathcal{D}^{(1)},\mathcal{D}^{(2)},\mathcal{D}^{(3)},$ for each of the functions are given for each coupling separately.

\begin{widetext}

\subsection{Even-polynomial GB coupling: $f(\vph)=\vph^{2}$}

\begin{eqnarray}
g_{tt}(r)&\approx & \mathcal{N}_{eve}^{(1)}/\mathcal{D}_{eve}^{(1)} \left(1-\frac{r_0}{r} \right) \,,
\\
\sqrt{g_{tt}(r) g_{rr}(r)}& \approx & \mathcal{N}_{eve}^{(2)}/\mathcal{D}_{eve}^{(2)}\,,
\\
e^{\vph (r)-\vph_\infty} &\approx & \mathcal{N}_{eve}^{(3)}/\mathcal{D}_{eve}^{(3)}\,,
\end{eqnarray}

where

\begin{eqnarray}
\mathcal{N}_{eve}^{(1)}&=&p^6 (0.0273395\, r_0^3-0.0273395\, r^2 r_0)+p^5 ( r^3-0.737869 \,r^2 r_0-0.0114842\, r r_0^2-0.273615\, r_0^3)+p^4 \,(-10.4088 \,r^3
\nonumber \\
&+&10.1241\, r^2 r_0+0.121014\, r r_0^2+1.15224\, r_0^3)+p^3 \,(29.8132\, r^3-34.4461\, r^2 r_0-0.419751\, r r_0^2-3.0735 \,r_0^3)
\nonumber \\
&+&p^2 (-7.36175\, r^3+25.9689\, r^2 r_0+0.540161\, r r_0^2+3.67187\, r_0^3)+p \,(-51.7172\, r^3+27.879 \,r^2 r_0-0.230379\, r r_0^2
\nonumber \\
&-&1.48583 \,r_0^3)+39.8966\, r^3-29.9903 \,r^2 r_0 \,,
\end{eqnarray}

\begin{eqnarray}
\mathcal{D}_{eve}^{(1)}&=& p^5 r^2 ( r- r_0)+p^4 r^2 (10.8289 \,r_0-10.4088\, r)+p^3 r^2 (29.8132 r-34.5843 r_0)+p^2 r^2 (24.7365 r_0-7.36175 r)
\nonumber\\
&+&p r^2 (28.8068 r_0-51.7172 r)+r^2 (39.8966 r-29.9903 r_0)\,,
\end{eqnarray}

\begin{eqnarray}
\mathcal{N}_{eve}^{(2)}&=& p^5 ( r^2- r r_0)+p^4 (5.97995\, r r_0-5.97995\, r^2)+p^3 (9.25097 r^2-9.03038\, r r_0+0.233666\, r_0^2)+p^2 (-13.5895 r^2
\nonumber\\
&+&12.1594 \,r r_0-0.516068 r_0^2)+p\, (22.3021 r^2-19.9922\, r r_0+0.283453 r_0^2)-13.2868\, r^2+12.1842\, r r_0 \,,
\end{eqnarray}

\begin{eqnarray}
\mathcal{D}_{eve}^{(2)}&=& p^5 r ( r- r_0)+p^4 r\, (5.97995 r_0-5.97995 r)+p^3 r (9.25097 r-9.03038 r_0)+p^2 r (12.1594 r_0-13.5895 r)
\nonumber\\
&+&p r \,(22.3021 r-19.9922 r_0)+r\, (12.1842 r_0-13.2868 r) \,,
\end{eqnarray}

\begin{eqnarray}
\mathcal{N}_{eve}^{(3)}&=& p^8 (r^2-1.06383 r r_0-0.210319 r_0^2)+p^7 (2.75803 r^2-1.60013 r r_0+0.823148 r_0^2)+p^6 (-4.75602 r^2+4.56272 r r_0
\nonumber\\
&-&1.0166 r_0^2)+p^5 (-9.42419 r^2+5.36685 r r_0+0.353077 r_0^2)+p^4 (7.2356 r^2-5.19214 r r_0+0.000836036 r_0^2)
\nonumber\\
&+&p^3 (4.81869 r^2-3.44584 r r_0+0.0311472 r_0^2)+p^2 (0.132709 r^2-0.0952131 r r_0+0.0000162571 r_0^2)
\nonumber\\
&+&p(0.000867808 r^2-0.000623499 r r_0)\,,
\end{eqnarray}
and

\begin{eqnarray}
\mathcal{D}_{eve}^{(3)}&=&  p^8 r (r-1.14693 r_0)+p^7 r (2.75803\, r-1.80427\, r_0)+p^6 r (5.00688 r_0-4.75602 r)+p^5 r (5.97025 r_0-9.42419\, r)
\nonumber\\
&+&p^4 r (7.2356 r-5.96539 r_0)+p^3 r (4.81869\, r-3.52478\, r_0)+p^2 r (0.132709 r-0.0963952 r_0)+p\, r (0.000867808 \,r
\nonumber\\
&-&0.000623594 \,r_0) \,.
\end{eqnarray}

\subsection{Even-polynomial GB coupling: $f(\vph)=\vph^{4}$}

\begin{eqnarray}
g_{tt}(r)&\approx & \mathcal{N}_{eve4}^{(1)}/\mathcal{D}_{eve4}^{(1)} \left(1-\frac{r_0}{r} \right) \,,
\\
\sqrt{g_{tt}(r) g_{rr}(r)}& \approx & \mathcal{N}_{eve4}^{(2)}/\mathcal{D}_{eve4}^{(2)}\,,
\\
e^{\vph (r)-\vph_\infty} &\approx & \mathcal{N}_{eve4}^{(3)}/\mathcal{D}_{eve4}^{(3)}\,,
\end{eqnarray}
where

\begin{eqnarray}
\mathcal{N}_{eve4}^{(1)}&=&p^8 (r^3-1.089 r^2 r_0+0.0889959 r_0^3)+p^7 (-1.87642 r^3+2.42973 r^2 r_0-0.0140959 r r_0^2-0.324535 r_0^3)
\nonumber\\
&+&p^6 \,(-6.91475\, r^3+6.18264\, r^2 r_0+0.0641805 \,r r_0^2+0.536171\, r_0^3)+p^5\, (13.6824\, r^3-14.7421 \,r^2 r_0
\nonumber\\
&-&0.0782894\, r r_0^2-0.400357 \,r_0^3)+p^4\, (-1.21904 \,r^3+3.58592\, \,r^2 \,r_0+0.0175305\, r r_0^2+0.0804516\, r_0^3)
\nonumber\\
&+&p^3 (-4.24836 r^3+3.42978 r^2 r_0+0.0100198\, r r_0^2+0.0134892\, r_0^3)+p^2 \,(-0.643342 r^3+0.437795\, r^2 r_0
\nonumber\\
&+&0.000722858\, r \,r_0^2+0.000594032\, r_0^3)+p \,(-0.02586 \,r^3+0.0142478 \,r^2 r_0)\,,
\end{eqnarray}

\begin{eqnarray}
\mathcal{D}_{eve4}^{(1)}&=&p^8 r^2 (r- r_0)+p^7 r^2 (2.0348 r_0-1.87642 r)+p^6 r^2 (6.59925 r_0-6.91475 r)+p^5 r^2 (13.6824 r-14.6863 r_0)
\nonumber\\
&+&p^4 r^2 (3.44811 r_0-1.21904 r)+p^3 r^2 (3.3963 r_0-4.24836 r)+p^2 r^2 (0.436086 r_0-0.643342 r)
\nonumber\\
&+&p r^2 (0.0142432 r_0-0.02586 r)\,,
\end{eqnarray}

\begin{eqnarray}
\mathcal{N}_{eve4}^{\,(2)}&=&p^5 \,(r^2-\,r\,r_0)+p^4 \,(-6.02549\,r^2+6.109\,r\,r_0+0.0865879\,r_0^2)+p^3 \,(7.52143\,r^2-8.05501\,r\,r_0-0.189351\,r_0^2)
\nonumber\\
&+&p^2 \,(0.812262\,r^2+0.0379139\,r\,r_0+0.100843\,r_0^2)+p \,(-3.41709\,r^2+3.02529\,r\,r_0+0.00237323\,r_0^2)
\nonumber\\
&-&0.0430078\,r^2+0.0337265\,r\,r_0
\,,
\end{eqnarray}

\begin{eqnarray}
\mathcal{D}_{eve4}^{(2)}&=&p^5 r (r-r_0)+p^4 r (6.109 r_0-6.02549 r)+p^3 r (7.52143 r-8.05501 r_0)+p^2 r (0.812262 r+0.0379139 r_0)
\nonumber\\
&+&p r (3.02529 r_0-3.41709 r)+r (0.0337265 r_0-0.0430078 r)
\,,
\end{eqnarray}

\begin{eqnarray}
\mathcal{N}_{eve4}^{(3)}&=& p^9 (0.00714286 r_0^2-0.00714286 r r_0)+p^8 ( r^2-0.830984 r r_0-0.169016 r_0^2)+p^7 (-1.71501 r^2+1.43139 r r_0
\nonumber\\
&+&0.287647 r_0^2)+p^6 (-2.8681 r^2+1.90683 r r_0-0.0388565 r_0^2)+p^5 (2.68777 r^2-2.38079 r r_0-0.239279 r_0^2)
\nonumber\\
&+&p^4 (3.08141 r^2-1.93109 r r_0-0.0512424 r_0^2)+p^3 (0.69013 r^2-0.376209 r r_0+0.000923929 r_0^2)
\nonumber\\
&+&p^2 (0.0167125 r^2-0.0085256 r r_0-0.0000264241 r_0^2)+p (0.00011179 r^2-0.0000470767 r r_0)
\,,
\end{eqnarray}

\begin{eqnarray}
\mathcal{D}_{eve4}^{(3)}&=& p^8 r (r- r_0)+p^7 r (1.71501 r_0-1.71501 r)+p^6 r (2.30399 r_0-2.8681 r)+p^5 r (2.68777 r-2.85841 r_0)
\nonumber\\
&+&p^4 r (3.08141 r-2.28967 r_0)+p^3 r (0.69013 r-0.413742 r_0)+p^2 r (0.0167125 r-0.00910661 r_0)
\nonumber\\
&+&p\, r (0.00011179 r-0.000048073 r_0)
\,.
\end{eqnarray}

\subsection{Odd-polynomial GB coupling: $f(\vph)=\vph^{3}$}

\begin{eqnarray}
g_{tt}(r)&\approx & \mathcal{N}_{odd}^{(1)}/\mathcal{D}_{odd}^{(1)} \left(1-\frac{r_0}{r} \right) \,,
\\
\sqrt{g_{tt}(r) g_{rr}(r)}& \approx & \mathcal{N}_{odd}^{(2)}/\mathcal{D}_{odd}^{(2)}\,,
\\
e^{\vph (r)-\vph_\infty} &\approx & \mathcal{N}_{odd}^{(3)}/\mathcal{D}_{odd}^{(3)}\,,
\end{eqnarray}
where

\begin{eqnarray}
\mathcal{N}_{odd}^{(1)}&=& p^8 (0.00537634 r^2 r_0-0.00537634 r_0^3)+p^7 (-0.0518836 r^2 r_0+0.00105719 r r_0^2+0.0529408 r_0^3)
\nonumber\\
&+&p^6 (r^3-0.86139 r^2 r_0-0.0104076 r r_0^2-0.149018 r_0^3)+p^5 (-3.96079 r^3+4.05559 r^2 r_0+0.029677 r r_0^2
\nonumber\\
&+&0.211035 r_0^3)+p^4 (3.89682 r^3-4.75407 r^2 r_0-0.0293161 r r_0^2-0.143895 r_0^3)+p^3 (0.572272 r^3+0.467441 r^2 r_0
\nonumber\\
&+&0.00542841 r r_0^2+0.0284754 r_0^3)+p^2 (-1.35561 r^3+1.05084 r^2 r_0+0.00352281 r r_0^2+0.00513373 r_0^3)
\nonumber\\
&+&p (-0.17693 r^3+0.113522 r^2 r_0+0.000048065 r r_0^2+0.000151556 r_0^3)-0.000822131 r^3 \,,
\end{eqnarray}

\begin{eqnarray}
\mathcal{D}_{odd}^{(1)}&=& p^6 r^2 (r-r_0)+p^5 r^2 (4.15743 r_0-3.96079 r)+p^4 r^2 (3.89682 r-4.71385 r_0)+p^3 r^2 (0.572272 r+0.426992 r_0)
\nonumber\\
&+&p^2 r^2 (1.04068 r_0-1.35561 r)+p r^2 (0.113474 r_0-0.17693 r)-0.000822131 r^3 \,,
\end{eqnarray}

\begin{eqnarray}
\mathcal{N}_{odd}^{(2)}&=& p^5 ( r^2- r r_0)+p^4 (-5.85214 r^2+5.95711 r r_0+0.108754 r_0^2)+p^3 (6.41291 r^2-7.08335 r r_0-0.238311 r_0^2)
\nonumber\\
&+&p^2 (2.69182 r^2-1.62168 r r_0+0.127734 r_0^2)+p\, (-4.37086 r^2+3.87431 r r_0+0.00238993 r_0^2)-0.0482146 r^2
\nonumber\\
&+&0.0388754 r r_0 \,,
\end{eqnarray}

\begin{eqnarray}
\mathcal{D}_{odd}^{(2)}&=& p^5 r (r-r_0)+p^4 r (5.95711 r_0-5.85214 r)+p^3 r (6.41291 r-7.08335 r_0)+p^2 r (2.69182 r-1.62168 r_0)
\nonumber\\
&+&p r \,(3.87431 r_0-4.37086 r)+r (0.0388754 r_0-0.0482146 r) \,,
\end{eqnarray}

\begin{eqnarray}
\mathcal{N}_{odd}^{(3)}&=& p^8 (0.00790584 r_0^2-0.00704225 r r_0)+p^7 (r^2-0.974721 r r_0-0.261618 r_0^2)+p^6 (2.50004 r^2-1.58963 r r_0
\nonumber\\
&+&0.231062 r_0^2)+p^5 (-0.277785 r^2+0.826815 r r_0+0.0230046 r_0^2)+p^4 (-3.50087 r^2+2.17273 r r_0+0.111646 r_0^2)
\nonumber\\
&+&p^3 (-1.36372 r^2+0.827451 r r_0+0.00569516 r_0^2)+p^2 (-0.0768515 r^2+0.0400392 r r_0-0.000361993 r_0^2)
\nonumber\\
&+&p (-0.00130153\, r^2+0.000634807\, r\, r_0)
\,,
\end{eqnarray}

\begin{eqnarray}
\mathcal{D}_{odd}^{(3)}&=& p^7 r (r-1.12263 r_0)+p^6 r (2.50004 r-1.95942 r_0)+p^5 r (0.945795 r_0-0.277785 r)+p^4 r (2.69322 r_0-3.50087 r)
\nonumber\\
&+&p^3 r (0.910352 r_0-1.36372 r)+p^2 r (0.0433891 r_0-0.0768515 r)+p r\, (0.000670351 r_0-0.00130153 r)
\,.
\end{eqnarray}

\subsection{Inverse-polynomial GB coupling $f(\vph)=\vph^{-1}$}

\begin{eqnarray}
g_{tt}(r)&\approx & \mathcal{N}_{inv}^{(1)}/\mathcal{D}_{inv}^{(1)} \left(1-\frac{r_0}{r} \right) \,,
\\
\sqrt{g_{tt}(r) g_{rr}(r)}& \approx & \mathcal{N}_{inv}^{(2)}/\mathcal{D}_{inv}^{(2)}\,,
\\
e^{\vph (r)-\vph_\infty} &\approx & \mathcal{N}_{inv}^{(3)}/\mathcal{D}_{inv}^{(3)}\,,
\end{eqnarray}
where

\begin{eqnarray}
\mathcal{N}_{inv}^{(1)}&=& p^8 (0.00630915 r^2 r_0-0.00630915 r_0^3)+p^7 (-0.270873 r^2 r_0+0.00158747 r r_0^2+0.272461 r_0^3)+p^6 ( r^3+1.68132 r^2 r_0
\nonumber\\
&-&0.0676888 r r_0^2-1.87658 r_0^3)+p^5 (-30.771 r^3+27.3744 r^2 r_0+0.65521 r r_0^2+3.28027 r_0^3)+p^4 (52.7707 r^3
\nonumber\\
&-&61.4833 r^2 r_0-0.744326 r r_0^2-0.405224 r_0^3)+p^3 (-0.028509 r^3+13.4581 r^2 r_0-0.310375 r r_0^2-2.09652 r_0^3)
\nonumber\\
&+&p^2 (-24.1726 r^3+25.4044 r^2 r_0+0.435715 r r_0^2+0.619006 r_0^3)+p\, (-0.865964 r^3-3.69793 r^2 r_0+0.0327455 r r_0^2
\nonumber\\
&+&0.11195 r_0^3)+0.604044 r^3-0.93029 r^2 r_0 \,,
\end{eqnarray}

\begin{eqnarray}
\mathcal{D}_{inv}^{(1)}&=& p^6 r^2 (r- r_0)+p^5 r^2 (31.0226 r_0-30.771 r)+p^4 r^2 (52.7707 r-60.4391 r_0)+p^3 r^2 (11.1237 r_0-0.028509 r)
\nonumber\\
&+&p^2 r^2 (25.2874 r_0-24.1726 r)+p r^2 (-0.865964 r-3.6373 r_0)+r^2 (0.604044 r-0.93029 r_0) \,,
\end{eqnarray}

\begin{eqnarray}
\mathcal{N}_{inv}^{(2)}&=& p^4 ( r^2-1.16356 r r_0-0.148553 r_0^2)+p^3 (-4.69714 r^2+6.11751 r r_0+0.733313 r_0^2)+p^2 (1.51773 r^2-5.7017 r r_0
\nonumber\\
&-&1.07164 r_0^2)+p\, (9.04644 r^2-4.16948 r r_0+0.487947 r_0^2)-7.0095\, r^2+5.05739\, r r_0 \,,
\end{eqnarray}

\begin{eqnarray}
\mathcal{D}_{inv}^{(2)}&=& p^4 r ( r-1.16356 r_0)+p^3 r (6.11751 r_0-4.69714 r)+p^2 r (1.51773 r-5.7017 r_0)+p r\, (9.04644 r-4.16948 r_0)
\nonumber\\
&+&r (5.05739 r_0-7.0095 r) \,,
\end{eqnarray}

\begin{eqnarray}
\mathcal{N}_{inv}^{(3)}&=& p^8 (-0.0140845 r r_0-0.136912 r_0^2)+p^7 (r^2+0.455729 r r_0-0.327295 r_0^2)+p^6 (-3.28574 r^2+1.52376 r r_0
\nonumber\\
&+&0.538745 r_0^2)+p^5 (0.397664 r^2-1.30549 r r_0+0.0866968 r_0^2)+p^4 (1.67086 r^2-0.71786 r r_0-0.115423 r_0^2)
\nonumber\\
&+&p^3 (0.381375 r^2-0.127462 r r_0-0.0145457 r_0^2)+p^2 (0.0258233 r^2-0.00845295 r r_0-0.000366578 r_0^2)
\nonumber\\
&+&p (0.00034624 r^2-0.000100545 r r_0)
\,,
\end{eqnarray}

\begin{eqnarray}
\mathcal{D}_{inv}^{(3)}&=& p^7 r (r+0.79726 r_0)+p^6 r (0.109978 r_0-3.28574 r)+p^5 r (0.397664 r-0.585227 r_0)+p^4 r (1.67086 r-0.367609 r_0)
\nonumber\\
&+&p^3 r (0.381375 r-0.0841323 r_0)+p^2 r (0.0258233 r-0.00668003 r_0)+p \,r\, (0.00034624 r-0.0000924255 r_0)
\,.
\end{eqnarray}

\subsection{Logarithmic GB coupling: $f(\vph)=\ln{(\vph)}$}

\begin{eqnarray}
g_{tt}(r)&\approx & \mathcal{N}_{log}^{(1)}/\mathcal{D}_{log}^{(1)} \left(1-\frac{r_0}{r} \right) \,,
\\
\sqrt{g_{tt}(r) g_{rr}(r)}& \approx & \mathcal{N}_{log}^{(2)}/\mathcal{D}_{log}^{(2)}\,,
\\
e^{\vph (r)-\vph_\infty} &\approx & \mathcal{N}_{log}^{(3)}/\mathcal{D}_{log}^{(3)}\,,
\end{eqnarray}
where

\begin{eqnarray}
\mathcal{N}_{log}^{(1)}&=& p^7 (0.0626084 r_0^3-0.0626084 r^2 r_0)+p^6 ( r^3-0.621986 r^2 r_0-0.378014 r_0^3)+p^5 (-4.81894 r^3+4.13961 r^2 r_0
\nonumber\\
&-&0.0186288 r r_0^2+0.65135 r_0^3)+p^4 (7.27136 r^3-6.76272 r^2 r_0+0.117437 r r_0^2-0.0407656 r_0^3)+p^3 (-1.24665 r^3
\nonumber\\
&+&0.285681 r^2 r_0-0.248593 r r_0^2-0.910056 r_0^3)+p^2 (-6.33774 r^3+8.6188 r^2 r_0+0.219636 r r_0^2+0.845321 r_0^3)
\nonumber\\
&+&p (5.38018 r^3-7.52226 r^2 r_0-0.0698525 r r_0^2-0.230411 r_0^3)-1.24797 r^3+1.92521 r^2 r_0 \,,
\end{eqnarray}

\begin{eqnarray}
\mathcal{D}_{log}^{(1)}&=& p^6 r^2 (r-r_0)+p^5 r^2 (4.81894 r_0-4.81894 r)+p^4 r^2 (7.27136 r-6.97382 r_0)+p^3 r^2 (-1.24665 r-0.266442 r_0)
\nonumber\\
&+&p^2 r^2 (9.14686 r_0-6.33774 r)+p r^2 (5.38018 r-7.65098 r_0)+r^2 (1.92521 r_0-1.24797 r) \,,
\end{eqnarray}

\begin{eqnarray}
\mathcal{N}_{log}^{(2)}&=& p^4 (r^2-0.593401 r r_0+0.394399 r_0^2)+p^3 (-7.9072 r^2+4.88145 r r_0-1.33585 r_0^2)+p^2 (20.3089 r^2-13.2461 r r_0
\nonumber\\
&+&1.50407 r_0^2)+p\, (-21.2712 r^2+14.5616 r r_0-0.562773 r_0^2)+7.88315 r^2-5.61692 r r_0 \,,
\end{eqnarray}

\begin{eqnarray}
\mathcal{D}_{log}^{(2)}&=& p^4 r (r-0.593401 r_0)+p^3 r (4.88145 r_0-7.9072 r)+p^2 r (20.3089 r-13.2461 r_0)+p\, r\, (14.5616 r_0-21.2712 r)
\nonumber\\
&+&r (7.88315 r-5.61692 r_0) \,,
\end{eqnarray}

\begin{eqnarray}
\mathcal{N}_{log}^{(3)}&=& p^8 (0.311804 r r_0-1.56376 r_0^2)+p^7 (r^2-0.552096 r r_0+5.09849 r_0^2)+p^6 (5.22019 r^2-1.79963 r r_0-5.43249 r_0^2)
\nonumber\\
&+&p^5 (-16.4094 r^2+4.36303 r r_0+1.52051 r_0^2)+p^4 (8.21674 r^2-1.77521 r r_0+0.254627 r_0^2)
\nonumber\\
&+&p^3 (2.36595 r^2-0.651287 r r_0+0.0795184 r_0^2)+p^2 (0.160339 r^2-0.0446282 r r_0+0.00490208 r_0^2)
\nonumber\\
&+&p (0.0026305 r^2-0.000740342 r r_0)
\,,
\end{eqnarray}

\begin{eqnarray}
\mathcal{D}_{log}^{(3)}&=& p^7 r (r-2.35737 r_0)+p^6 r (5.22019 r+2.40632 r_0)+p^5 r (4.7997 r_0-16.4094 r)+p^4 r (8.21674 r-4.25309 r_0)
\nonumber\\
&+&p^3 r (2.36595 r-0.95373 r_0)+p^2 r (0.160339 r-0.0538552 r_0)+p r (0.0026305 r-0.000823782 r_0)
\,.
\end{eqnarray}

\end{widetext}

\section{Analytical expressions for the higher-order CFA coefficients up to fourth order}
\label{App B}

\subsection{Even-polynomial GB coupling: $f(\vph)=\vph^2$}

\beq
a_3= \frac{-\frac{13 p^2}{287}+p-\frac{37}{125}}{\frac{177}{991}-\frac{56 p}{313}}\,,
\label{a3 even}
\eeq

\beq
a_4= \frac{-\frac{463 p^4}{1015}+p^3-\frac{233 p^2}{338}+\frac{47 p}{363}-\frac{1}{189}}{\frac{p}{353}-\frac{1}{350}}\,,
\label{a4 even}
\eeq

\beq
b_3=\frac{-\frac{32 p^2}{89}+p-\frac{79}{156}}{\frac{8 p^2}{131}-\frac{56 p}{375}+\frac{65}{737}}\,,
\label{b3 even}
\eeq

\beq
b_4=\frac{p^2-\frac{37 p}{90}+\frac{17}{366}}{\frac{10}{303}-\frac{43 p}{1375}}\,,
\label{b4 even}
\eeq

\beq
f_3 = \frac{-\frac{18 p^2}{73}+p-\frac{59}{187}}{\frac{37}{133}-\frac{28 p}{103}} \,,
\label{f3 even}
\eeq

\beq
f_4 = \frac{-\frac{54 p^4}{115}+p^3-\frac{50 p^2}{77}+\frac{10 p}{97}-\frac{1}{477}}{\frac{p}{463}-\frac{1}{456}} \,.
\label{f4 even}
\eeq

\subsection{Even-polynomial GB coupling: $f(\vph)=\vph^4$}

\beq
a_3= \frac{-\frac{103 p^3}{265}+p^2-\frac{41 p}{376}-\frac{1}{476}}{\frac{74 p^3}{391}-\frac{449 p^2}{914}+\frac{65 p}{214}-\frac{1}{530}}\,,
\label{a3 even quartic}
\eeq

\beq
a_4= \frac{-\frac{123 p^4}{298}+p^3-\frac{502 p^2}{575}+\frac{93 p}{293}+\frac{39}{413}}{\frac{5}{262}-\frac{7 p}{367}}\,,
\label{a4 even quartic}
\eeq

\beq
b_3=\frac{p^2-\frac{37 p}{317}-\frac{97}{313}}{\frac{7 p^2}{291}-\frac{52 p}{339}+\frac{46}{357}}\,,
\label{b3 even quartic}
\eeq

\beq
b_4= \frac{-\frac{191 p^3}{244}+p^2-\frac{113 p}{559}+\frac{6}{419}}{\frac{2}{271}-\frac{p}{134}}\,,
\label{b4 even quartic}
\eeq

\beq
f_3 =  \frac{-\frac{45 p^2}{167}+p-\frac{3}{47}}{\frac{13}{41}-\frac{13 p}{42}}\,,
\label{f3 even quartic}
\eeq

\beq
f_4 = \frac{-\frac{27 p^4}{70}+p^3-\frac{62 p^2}{63}+\frac{46 p}{105}-\frac{1}{74}}{\frac{1}{100}-\frac{p}{100}}\,.
\label{f4 even quartic}
\eeq

\subsection{Odd-polynomial GB coupling: $f(\vph)=\vph^3$}

\beq
a_3=\frac{\frac{127 p^3}{188}-\frac{257 p^2}{270}+p-\frac{34}{213}}{\frac{35}{253}-\frac{49 p}{354}}\,,
\label{a3 odd}
\eeq

\beq
a_4=\frac{-\frac{70 p^3}{131}+p^2-\frac{167 p}{301}-\frac{35}{367}}{\frac{28 p}{1009}-\frac{8}{287}}\,,
\label{a4 odd}
\eeq

\beq
b_3 = \frac{p^2-\frac{7 p}{260}-\frac{74}{197}}{\frac{5 p^2}{296}-\frac{29 p}{184}+\frac{37}{264}} \,,
\label{b3 odd}
\eeq

\beq
b_4 = \frac{p^2-\frac{409 p}{1002}+\frac{15}{377}}{\frac{5}{144}-\frac{11 p}{331}} \,,
\label{b4 odd}
\eeq

\beq
f_3 = \frac{-\frac{19 p^2}{78}+p-\frac{21}{230}}{\frac{23}{64}-\frac{20 p}{57}} \,,
\label{f3 odd}
\eeq

\beq
f_4 = \frac{-\frac{110 p^4}{283}+p^3-\frac{62 p^2}{65}+\frac{50 p}{129}-\frac{1}{140}}{\frac{1}{126}-\frac{p}{126}}\,.
\label{f4 odd}
\eeq

\subsection{Inverse-polynomial GB coupling: $f(\vph)=\vph^{-1}$}

\beq
a_3 = \frac{\frac{99 p^3}{817}-\frac{70 p^2}{299}+p+\frac{37}{524}}{-\frac{106 p^2}{207}+\frac{71 p}{144}+\frac{5}{276}} \,,
\label{a3 inv}
\eeq

\beq
a_4 = \frac{-\frac{110 p^3}{197}+p^2-\frac{13 p}{94}-\frac{2}{181}}{\frac{21}{185}-\frac{43 p}{379}} \,,
\label{a4 inv}
\eeq

\beq
b_3 = \frac{-\frac{63 p^2}{193}+p-\frac{101}{247}}{\frac{37 p^2}{211}-\frac{16 p}{35}+\frac{69}{245}} \,,
\label{b3 inv}
\eeq

\beq
b_4 = \frac{-\frac{266 p^3}{353}+p^2-\frac{77 p}{325}+\frac{1}{90}}{\frac{1}{57}-\frac{6 p}{337}} \,,
\label{b4 inv}
\eeq

\beq
f_3 = \frac{-\frac{85 p^3}{117}+p^2-\frac{9 p}{116}-\frac{1}{235}}{-\frac{55 p^2}{439}+\frac{7 p}{59}+\frac{1}{106}} \,,
\label{f3 inv}
\eeq

\beq
f_4 = \frac{p^3-\frac{55 p^2}{71}+\frac{58 p}{209}}{-\frac{31 p^3}{67}+\frac{83 p^2}{117}-\frac{31 p}{104}+\frac{5}{98}} \,.
\label{f4 inv}
\eeq

\subsection{Logarithmic GB coupling: $f(\vph)=\log{(\vph)}$}

\beq
a_3 = \frac{p^3-\frac{1424 p^2}{2197}-\frac{239 p}{728}-\frac{49}{886}}{-\frac{8 p^3}{595}+\frac{71 p^2}{433}-\frac{99 p}{743}-\frac{5}{343}} \,,
\label{a3 log}
\eeq

\beq
a_4 = \frac{-\frac{637 p^3}{906}+p^2-\frac{132 p}{335}+\frac{15}{649}}{-\frac{98 p^2}{647}+\frac{487 p}{1523}-\frac{109}{644}} \,,
\label{a4 log}
\eeq

\beq
b_3 = \frac{-\frac{109 p^2}{181}+p-\frac{63}{188}}{\frac{8 p^2}{43}-\frac{97 p}{232}+\frac{55}{237}} \,,
\label{b3 log}
\eeq

\beq
b_4 = \frac{-\frac{93 p^3}{119}+p^2-\frac{24 p}{109}+\frac{7}{559}}{\frac{10}{287}-\frac{6 p}{169}} \,,
\label{b4 log}
\eeq

\beq
f_3 = \frac{-\frac{203 p^3}{347}+p^2-\frac{5 p}{64}-\frac{3}{361}}{-\frac{347 p^2}{523}+\frac{71 p}{108}+\frac{2}{95}} \,,
\label{f3 log}
\eeq

\beq
f_4 = \frac{-\frac{143 p^3}{331}+p^2-\frac{108 p}{161}}{\frac{22 p}{293}-\frac{17}{225}} \,.
\label{f4 log}
\eeq

\section{Analytical expressions for the photon-sphere radii and the black-hole shadows}
\label{App C}

\begin{figure*}
\includegraphics[angle=0.0,width=0.5\linewidth]{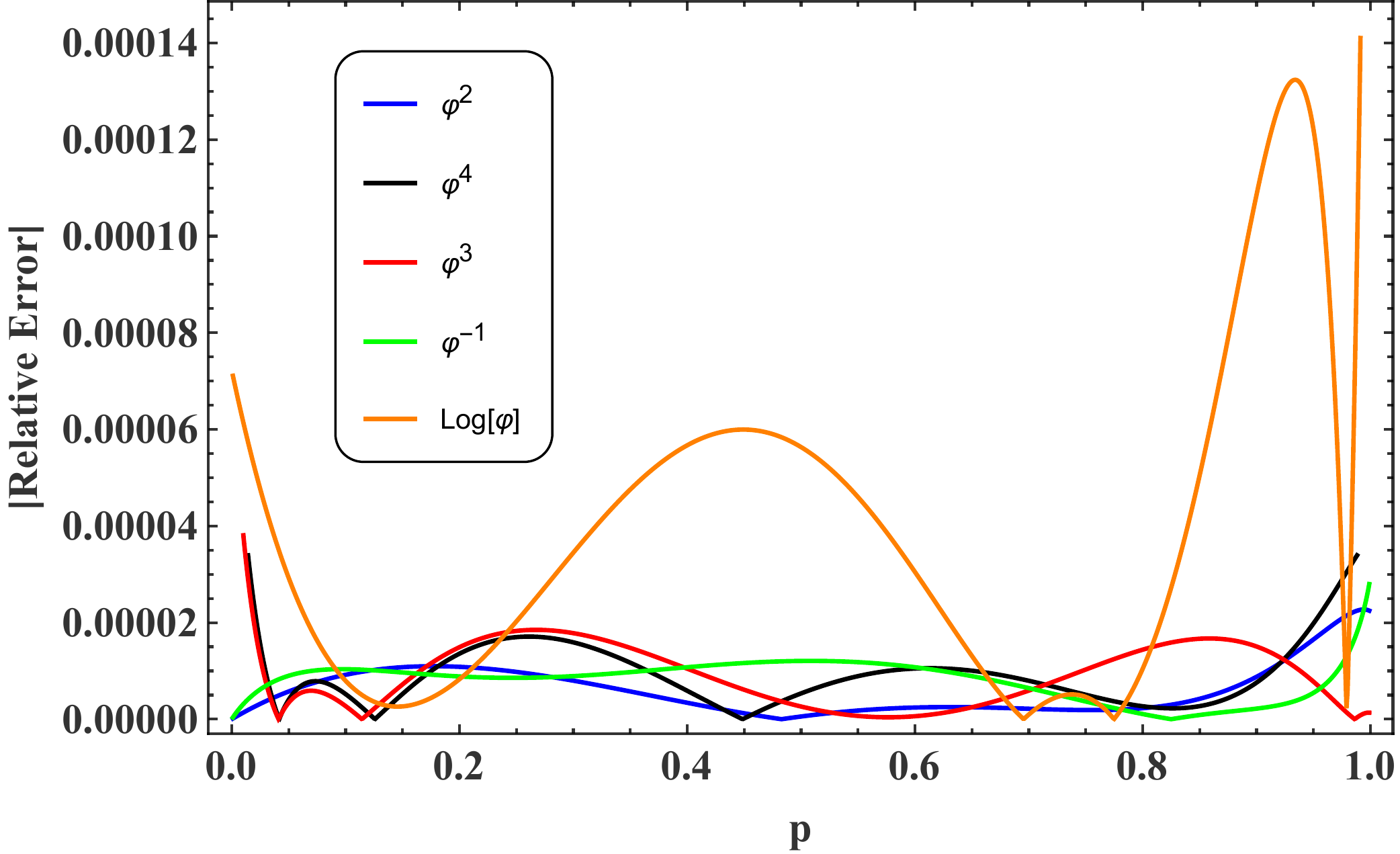}\includegraphics[angle=0.0,width=0.5\linewidth]{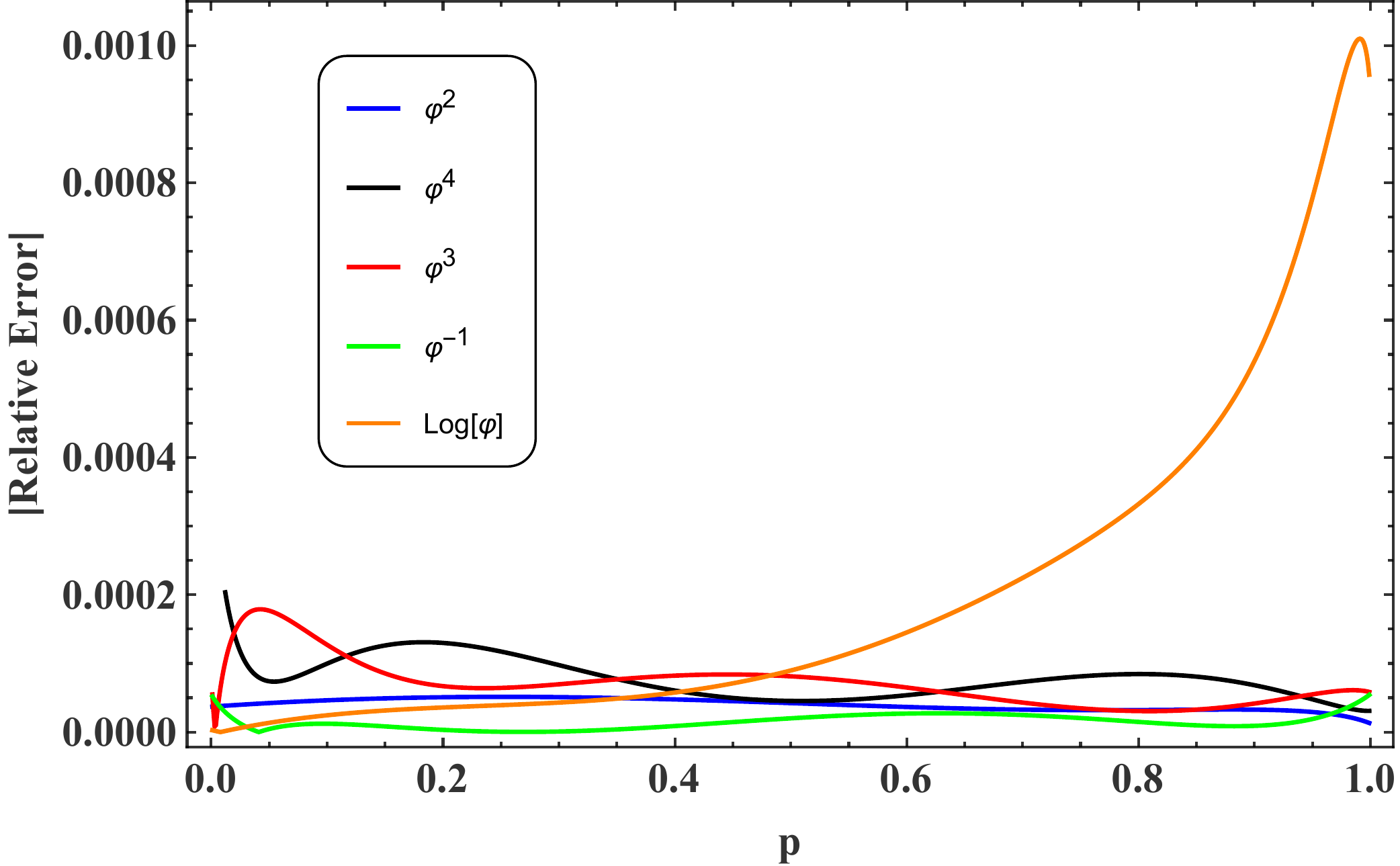}
\caption{The absolute relative error between numerical values and the analytical expressions of Eqs.\eqref{AppC eq1}-\eqref{AppC eq8} that have been obtained by fitting the accurate numerical values for the photon-sphere radius (left) and the black-hole shadow (right).}
\label{fig: RE_rph_and_shadow_anal_all}
\end{figure*}

Here we present approximate analytical expressions for the radius of the photon sphere and the black-hole shadow for the four GB couplings that we have considered in this work.

Notice that in order to obtain these analytical expressions no approximate expression for the metric functions has been involved. Instead, we employed only the accurate numerical solution for $g_{tt}(r)$ aiming to get the most accurate results.

For various values of $p$ we computed the corresponding values of $r_{ph}$ and $R_{sh}$ and in turn we performed a fitting of the collected data. Eventually, as any fitting procedure unavoidably introduces some error we have also included Fig.~\ref{fig: RE_rph_and_shadow_anal_all} to quantify the accuracy of the fitting of the numerical data at each value of the dimensionless parameter $p$.

\subsection{Even-polynomial GB coupling: $f(\vph)=\vph^2$}

\beq
r_{ph}= \frac{-\frac{5 p^2}{292}-\frac{366 p}{571}+1}{\frac{2}{3}-\frac{199 p}{448}}\,,
\label{AppC eq1}
\eeq

\beq
R_{sh}= \frac{-\frac{3 p^2}{292}-\frac{221 p}{617}+1}{\frac{102}{265}-\frac{55 p}{373}}\,.
\eeq

\subsection{Even-polynomial GB coupling: $f(\vph)=\vph^4$}

\beq
r_{ph}= \frac{\frac{3 p^3}{163}+p^2-\frac{265 p}{293}-\frac{236}{247}}{\frac{59 p^2}{88}-\frac{103 p}{177}-\frac{100}{157}}\,,
\eeq

\beq
R_{sh}= \frac{-\frac{4 p^3}{367}-\frac{164 p^2}{405}+p+\frac{77}{236}}{-\frac{162 p^2}{995}+\frac{127 p}{335}+\frac{28}{223}}
\,.
\eeq

\subsection{Odd-polynomial GB coupling: $f(\vph)=\vph^3$}

\beq
r_{ph}= \frac{\frac{8 p^3}{353}+p^2-\frac{293 p}{310}-\frac{249}{274}}{\frac{25 p^2}{37}-\frac{221 p}{362}-\frac{146}{241}}\,,
\eeq

\beq
R_{sh}= \frac{-\frac{3 p^3}{245}-\frac{311 p^2}{777}+p+\frac{183}{575}}{-\frac{20 p^2}{123}+\frac{115 p}{303}+\frac{55}{449}}\,.
\eeq

\subsection{Inverse-polynomial GB coupling: $f(\vph)=\vph^{-1}$}

\beq
r_{ph}= \frac{-\frac{3 p^3}{175}-\frac{61 p^2}{250}-\frac{146 p}{557}+1}{-\frac{73 p^2}{445}-\frac{41 p}{206}+\frac{2}{3}}\,,
\eeq

\beq
R_{sh}= \frac{-\frac{9 p^3}{407}-\frac{61 p^2}{174}+\frac{106 p}{157}+1}{-\frac{67 p^2}{454}+\frac{40 p}{171}+\frac{107}{278}}\,.
\eeq

\subsection{Logarithmic GB coupling: $f(\vph)=\ln{(\vph)}$}

\beq
r_{ph}= \frac{-\frac{3 p^3}{206}-\frac{131 p^2}{432}-\frac{41 p}{229}+1}{-\frac{81 p^2}{389}-\frac{181 p}{1266}+\frac{6925}{10388}}\,,
\eeq

\beq
R_{sh}= \frac{-\frac{2 p^3}{259}-\frac{89 p^2}{366}+p-\frac{523}{637}}{-\frac{37 p^2}{320}+\frac{172 p}{423}-\frac{73}{231}}\,.
\label{AppC eq8}
\eeq

\bibliography{References}{}
\bibliographystyle{utphys}
\end{document}